\documentclass[prc,twocolumn,showpacs]{revtex4}

\usepackage{graphicx}
\usepackage{dcolumn}
\usepackage{bm}

\begin{document}

\title{Gamow-Teller response in deformed even and odd neutron-rich Zr and Mo isotopes}

\author{P. Sarriguren$^{1}$}
\email{p.sarriguren@csic.es}
\author{A. Algora$^{2,3}$}
\author {J. Pereira$^{4,5}$}
\affiliation{
$^{1}$ Instituto de Estructura de la Materia, IEM-CSIC, Serrano
123, E-28006 Madrid, Spain \\
$^{2}$ Instituto de F\'\i sica Corpuscular , CSIC-Universitat de Valencia, E-46071 
Valencia, Spain \\
$^{3}$ Institute of Nuclear Research of the Hungarian Academy of Sciences, 
4026 Debrecen, Hungary \\
$^{4}$ National Superconducting Cyclotron Laboratory, MSU,
East Lansing, MI-48823, US\\
$^{5}$ Joint Institute for Nuclear Astrophysics (JINA), MSU, East Lansing, 
MI-48823, US}


\date{\today}

\begin{abstract}

$\beta$-decay properties of neutron-rich Zr and Mo isotopes are 
investigated within a microscopic theoretical approach based on 
the proton-neutron quasiparticle random-phase approximation. The 
underlying mean field is described self-consistently from deformed 
Skyrme Hartree-Fock calculations with pairing correlations.
Residual separable particle-hole and particle-particle forces
are also included in the formalism. The structural evolution in 
these isotopic chains including both even and odd isotopes is 
analyzed in terms of the equilibrium deformed shapes. Gamow-Teller 
strength distributions, $\beta$-decay half-lives, and $\beta$-delayed 
neutron-emission probabilities are studied, stressing their relevance 
to describe the path of the nucleosynthesis rapid neutron capture 
process.

\end{abstract}


\maketitle

\section{Introduction}

Neutron-rich nuclei in the mass region $A\sim 110-120$ have attracted
considerable attention from both theoretical and experimental sides in 
the last decades. They are interesting in many respects. First, the mass 
region is characterized by rapid structural changes in the ground state 
and low-lying collective excited states (see e.g. \cite{wood92,heyde11} 
for a general review). Relativistic \cite{xiang_12,mei_12} and 
nonrelativistic \cite{bonche85,bender08,bender09,rayner1,rayner2} 
studies of the nuclear structural evolution in this mass region show 
that the equilibrium shape of the nucleus changes rapidly with the 
number of nucleons and shape coexistence is present with competing 
spherical, axially symmetric prolate and oblate, as well as triaxial 
shapes at close energies. These features are supported by spectroscopic 
observations \cite{sumikama_11} including $2^+$ lifetime measurements
\cite{mach,goodin,urban} and quadrupole moments for rotational 
bands \cite{urban}, as well as by laser spectroscopy measurements
on Zr \cite{campbell} and Mo \cite{charlwood} isotopes.

In addition, this mass region is also involved in the astrophysical 
rapid neutron capture process (r process). The r process is considered 
to be one of the main nucleosynthesis mechanisms leading to the 
production of heavy neutron-rich nuclei and for the existence  of 
about half of the nuclei heavier than iron \cite{bbhf,cowan91}.
The r-process nucleosynthesis involves many neutron-rich unstable 
isotopes, whose masses and $\beta$-decay properties, including
$\beta$-decay half-lives ($T_{1/2}$) and $\beta$-delayed neutron-emission 
probabilities ($P_n$), are crucial quantities to understand the possible 
r-process paths, the isotopic abundances, and the time scales of the 
process \cite{cowan91,kra93}. Although much progress is being done 
recently on masses (see for example the Jyv\"askyl\"a mass database 
\cite{jyvaskyla}) and half-lives \cite{pereira,nishimura_11} measurements, 
unfortunately most of the nuclear properties of relevance for the 
r process are experimentally unknown due to their extremely low production 
yields in the laboratory. Therefore, reliable nuclear physics models are 
needed to interpret the astrophysical observations and to model and 
simulate properly the r process. For example, the agreement with the 
observed r-process abundances in this mass region is manifestly improved
\cite{pfeiffer01,sun08} when using nuclear structure models that include 
a shell quenching effect at $N=82$ \cite{doba96,ETFSIQ}.

The quasiparticle random-phase approximation (QRPA) has been shown to 
be a suitable model to deal with medium-mass open-shell nuclei.
QRPA calculations for neutron-rich nuclei have been performed
within different approaches, such as spherical formalisms based
on Hartree-Fock-Bogoliubov (HFB) mean field \cite{engel}, on
continuum spherical QRPA with density functionals \cite{borzov3}, and 
on relativistic mean field approaches \cite{niksic2005}. However, the 
mass region of concern here requires nuclear deformation as a relevant 
degree of freedom to characterize the nuclear structure involved in 
the calculation of the $\beta$-strength functions. The deformed QRPA 
formalism was developed in Refs. 
\cite{moller1,moller2,moller3,moller08,homma,hir1,hir2}, where 
phenomenological mean fields based on Nilsson or Woods-Saxon potentials 
were used as a starting basis. A Tamm-Dancoff approximation with Sk3 
interaction was also implemented in Ref. \cite{hamamoto}.

In this work we investigate the decay properties of both even and odd
neutron-rich  Zr and Mo isotopes within a deformed proton-neutron QRPA
based on a self-consistent Hartree-Fock (HF) mean field formalism with 
Skyrme interactions and pairing correlations in BCS approximation. 
Residual spin-isospin interactions are also included in the 
particle-hole and particle-particle channels \cite{sarri1,sarri2}.
This work was started in Ref. \cite{sarri_pere}, where the $\beta$-decay
properties of the even-even isotopes $^{100-110}$Zr and $^{104-114}$Mo
were studied. In the present work we extend that study in several aspects. 
First, we repeat the calculations of $T_{1/2}$ and $P_n$ for the same 
isotopes, but using new or updated measurements for the $Q_\beta$ and 
$S_n$ values \cite{audi12}.
Second, we extend the calculations to include odd-$A$ isotopes, as
well as more exotic isotopes, namely $^{112-116}$Zr and $^{116-120}$Mo.
This study is justified by a renewed experimental interest to measure
the half-lives of heavier and more exotic nuclei in this mass region 
at RIKEN. This study will contribute with
theoretical support to the understanding of these future measurements. 

The paper is organized as follows. In Sec. \ref{sec2} we present a 
review  of the theoretical formalism used. Sec. \ref{results} 
contains the results obtained for the potential energy curves (PEC), 
Gamow-Teller (GT) strength distributions, $\beta$-decay 
half-lives, and $\beta$-delayed neutron-emission probabilities.
Sec. IV summarizes the main conclusions.

\section{Theoretical Formalism}
\label{sec2}

In this section we show briefly the theoretical framework used in this
paper to describe the $\beta$-decay properties in Zr and Mo neutron-rich
isotopes. More details of the formalism can be found in 
Refs. \cite{sarri1,sarri2}.
The method consists of a self-consistent formalism based on a deformed
Hartree-Fock mean field obtained with Skyrme interactions including
pairing correlations. The single-particle energies, wave functions, 
and occupation probabilities are generated from this mean field.
In this work we have chosen the Skyrme force SLy4 \cite{sly4} as a
representative of the Skyrme forces. The protocol to derive this 
interaction includes information on the equation of state of 
pure neutron matter obtained from realistic forces that produce 
better isotopic properties. SLy4 is one of the more successful 
Skyrme forces and has been extensively studied in the last years 
\cite{bender08,bender09,stoitsov}.

The solution of the HF equation is found by using the formalism 
developed in Ref. \cite{vautherin}, assuming time reversal and axial 
symmetry. The single-particle wave functions are expanded in terms 
of the eigenstates of an axially symmetric harmonic oscillator in 
cylindrical coordinates, using twelve major shells. The method also 
includes pairing between like nucleons in BCS approximation with 
fixed gap parameters for protons and neutrons, which are determined
phenomenologically from the odd-even mass differences through a
symmetric five term formula involving the experimental binding
energies \cite{audi12} when available. In those cases where 
experimental information for masses is still not available, we have 
used the same pairing gaps as the closer isotopes measured. 
The pairing gaps for protons ($\Delta_p$) and neutrons ($\Delta_n$)
obtained in this way are roughly around 1 MeV.

The PECs are analyzed as a function of the quadrupole deformation 
parameter $\beta$. For that purpose, constrained 
HF calculations are performed with a quadratic constraint 
\cite{constraint}. The HF energy is minimized under the constraint 
of keeping fixed the nuclear deformation. Calculations for GT 
strengths are performed subsequently for the equilibrium shapes of 
each nucleus, that is, for the solutions, in general deformed, for 
which minima are obtained in the energy curves. Since decays 
connecting different shapes are disfavored, similar shapes are 
assumed for the ground state of the parent nucleus and for all
populated states in the daughter nucleus. The validity of this 
assumption was discussed for example in Refs. \cite{moller1,homma}. 

To describe GT transitions, a spin-isospin residual interaction is
added to the mean field and treated in a deformed proton-neutron QRPA
\cite{moller1,moller2,moller3,moller08,homma,hir1,hir2,hamamoto,sarri1,sarri2}.
This interaction contains two parts, particle-hole ($ph$) and 
particle-particle ($pp$). The interaction in the $ph$ channel is 
responsible for the position and structure of the GT resonance 
\cite{sarri1,sarri2,homma} and it can be derived consistently from 
the same Skyrme interaction used to generate the mean field, through 
the second derivatives of the energy density functional with respect 
to the one-body densities. The $ph$ residual interaction is finally 
expressed in a separable form by averaging the Landau-Migdal 
resulting force over the nuclear volume, as explained in Refs. 
\cite{sarri1,sarri2}. By taking separable GT forces, the energy 
eigenvalue problem reduces to find the roots of an algebraic 
equation. The $pp$ component is a neutron-proton pairing force in 
the $J^\pi=1^+$  coupling channel, which is also introduced as a 
separable force \cite{hir1,hir2,sarri2}. This strength is usually 
fitted to reproduce globally the experimental half-lives. Various 
attempts have been done in the past to fix this strength \cite{homma}, 
arriving to expressions that depend on the model used to describe the 
mean field, Nilsson model in the above reference.

In previous works \cite{sarri1,sarri2,sarri3,sarri4,sarri_wp} we 
studied the sensitivity of the GT strength distributions to the 
various ingredients contributing to the deformed QRPA calculations, 
namely to the nucleon-nucleon effective force, to pairing correlations,
and to residual interactions. We found different sensitivities to 
them. In this work, all of these ingredients have been fixed to the 
most reasonable choices found previously. In particular we use the 
coupling strengths  $\chi ^{ph}_{GT}=0.15$ MeV and 
$\kappa ^{pp}_{GT} = 0.03$ MeV for the $ph$ and $pp$ channels, 
respectively.

The technical details to solve the QRPA equations have been described 
in Refs. \cite{hir1,hir2,sarri1}. Here we only mention that, because 
of the use of separable residual forces, the solutions of the QRPA 
equations are found by solving first a dispersion relation, which is 
an algebraic equation of fourth order in the excitation energy $\omega$. 
Then, for each value of the energy, the GT transition amplitudes in 
the intrinsic frame connecting the ground state $| 0^+\rangle $ of an 
even-even nucleus to one phonon states in the daughter nucleus 
$|\omega_K \rangle \, (K=0,1) $ are found to be

\begin{equation}
\left\langle \omega _K | \sigma _K t^{\pm} | 0 \right\rangle =
\mp M^{\omega _K}_\pm \, ,
\label{intrinsic}
\end{equation}
where $t^+ |\pi \rangle =|\nu \rangle,\, t^- |\nu \rangle =|\pi \rangle$ and
\begin{eqnarray}
M_{-}^{\omega _{K}}&=&\sum_{\pi\nu}\left( q_{\pi\nu}X_{\pi
\nu}^{\omega _{K}}+ \tilde{q}_{\pi\nu}Y_{\pi\nu}^{\omega _{K}}
\right) , \\
M_{+}^{\omega _{K}}&=&\sum_{\pi\nu}\left(
\tilde{q}_{\pi\nu} X_{\pi\nu}^{\omega _{K}}+
q_{\pi\nu}Y_{\pi\nu}^{\omega _{K}}\right) \, ,
\end{eqnarray}
with
\begin{equation}
\tilde{q}_{\pi\nu}=u_{\nu}v_{\pi}\Sigma _{K}^{\nu\pi },\ \ \
q_{\pi\nu}=v_{\nu}u_{\pi}\Sigma _{K}^{\nu\pi},
\label{qs}
\end{equation}
in terms of the occupation amplitudes for neutrons and protons $v_{\nu,\pi}$   
($u^2_{\nu,\pi}=1-v^2_{\nu,\pi}$) and the matrix elements of the spin operator, 
$\Sigma _{K}^{\nu\pi}=\left\langle \nu\left| \sigma _{K}\right| 
\pi\right\rangle $,
connecting proton and neutron single-particle states, as they come out 
from the HF+BCS calculation. $X_{\pi\nu}^{\omega _{K}}$ and 
$Y_{\pi\nu}^{\omega _{K}}$ are the forward and backward amplitudes of the 
QRPA phonon operator, respectively. 

Once the intrinsic amplitudes in Eq. (\ref{intrinsic}) are calculated, 
the GT strength $B_{\omega}(GT^\pm)$ in the laboratory system for a 
transition  $I_iK_i (0^+0) \rightarrow I_fK_f (1^+K)$ can be obtained as
\begin{eqnarray}
B_{\omega}(GT^\pm )& =& \sum_{\omega_{K}} \left[ \left\langle \omega_{K=0}
\left| \sigma_0t^\pm \right| 0 \right\rangle ^2 \delta (\omega_{K=0}-
\omega ) \right.  \nonumber  \\
&& \left. + 2 \left\langle \omega_{K=1} \left| \sigma_1t^\pm \right|
0 \right\rangle ^2 \delta (\omega_{K=1}-\omega ) \right] \, ,
\label{bgt}
\end{eqnarray}
in $[g_A^2/4\pi]$ units. To obtain this expression, the initial and
final states in the laboratory frame have been expressed in terms of
the intrinsic states using the Bohr-Mottelson factorization \cite{bm}.

When the parent nucleus has an odd nucleon, the ground state can be 
expressed as a one-quasiparticle (1qp) state in which the odd nucleon 
occupies the single-particle orbit of lowest energy. Then two types 
of transitions are possible. One type is due to phonon excitations 
in which the odd nucleon acts only as a spectator. These are 
three-quasiparticle (3qp) states. In the intrinsic frame, the 
transition amplitudes are in this case basically the same as in 
the even-even case in Eq. (\ref{intrinsic}), but with the blocked 
spectator excluded from the calculation. The other type of transitions 
are those involving the odd nucleon state (1qp), which are treated 
by taking into account phonon correlations in the quasiparticle 
transitions in first-order perturbation. The transition amplitudes 
for the correlated states can be found in Ref. \cite{hir2,sarri4}.

Concerning the excitation energy of the daughter nuclei to
which we refer all the GT strength distributions in this paper,
we have to distinguish again between the case of even-even
and odd-$A$ parents. In the case of even-even systems, the
excitation energy of the 2qp states is simply given by

\begin{equation}
E_{\mbox{\scriptsize{ex}}\, [(Z,N)\rightarrow (Z+1,N-1)]}=\omega -E_{\pi_0} -
E_{\nu_0} \, , 
\label{eexeven}
\end{equation}
where $E_{\pi_0}$ and $E_{\nu_0}$ are the lowest quasiparticle energies 
for protons and neutrons, respectively. In the case of an odd-$A$ 
nucleus we have to deal with 1qp and 3qp transitions. For Zr and Mo
isotopes we have an odd-neutron parent decaying into an odd-proton 
daughter.  The excitation energies for 1qp transitions are

\begin{equation}
E_{\mbox{\scriptsize{ex,1qp}}\, [(Z,N+1)\rightarrow (Z+1,N)]}=E_\pi-E_{\pi_0} \, . 
\label{eex1qp}
\end{equation}
On the other hand, in the 3qp case where the unpaired neutron
acts as a spectator, the excitation energy with respect to
the ground state of the daughter nucleus is

\begin{equation}
E_{\mbox{\scriptsize{ex,3qp}}\, [(Z,N+1)\rightarrow (Z+1,N)]}=
\omega +E_{\nu,\mbox{\scriptsize{spect}}}-E_{\pi_0} \, . 
\label{eex3qp}
\end{equation}
This implies that the lowest excitation energy of 3qp type
is of the order of twice the neutron pairing gap. Therefore,
all the strength contained in the low-excitation-energy
region below typically 2-3 MeV in the odd-A nuclei studied
in this paper, corresponds to 1qp transitions.

The $\beta$-decay half-life is obtained by summing all the allowed
transition strengths to states in the daughter nucleus with
excitation energies lying below the corresponding $Q$-energy,
$Q_\beta\equiv Q_{\beta^-}= M(A,Z)-M(A,Z+1)-m_e $, written in terms of 
the nuclear masses $M(A,Z)$ and the electron mass ($m_e$), and
weighted with the phase space factors $f(Z,Q_{\beta}-E_{ex})$,

\begin{equation}
T_{1/2}^{-1}=\frac{\left( g_{A}/g_{V}\right) _{\rm eff} ^{2}}{D}
\sum_{0 < E_{ex} < Q_\beta}f\left( Z,Q_{\beta}-E_{ex} \right) B(GT,E_{ex}) \, ,
 \label{t12}
\end{equation}
with $D=6200$~s and $(g_A/g_V)_{\rm eff}=0.77(g_A/g_V)_{\rm free}$,
where 0.77 is a standard quenching factor.  The bare results can be 
recovered by scaling the results in this paper for $B(GT)$ and 
$T_{1/2}$ with the square of this quenching factor.

The Fermi integral $f(Z,Q_{\beta}-E_{ex})$ is computed numerically for each value 
of the energy including screening and finite size effects, as explained 
in Ref. \cite{gove},

\begin{equation}
f^{\beta^\pm} (Z, W_0) = \int^{W_0}_1 p W (W_0 - W)^2 \lambda^\pm(Z,W)
{\rm d}W\, ,
\end{equation}
with

\begin{equation}
\lambda^\pm(Z,W) = 2(1+\gamma) (2pR)^{-2(1-\gamma)} e^{\mp\pi y}
\frac{|\Gamma (\gamma+iy)|^2}{[\Gamma (2\gamma+1)]^2}\, ,
\end{equation}
where $\gamma=\sqrt{1-(\alpha Z)^2}$ ; $y=\alpha ZW/p$ ; $\alpha$ is
the fine structure constant and $R$ the nuclear radius. $W$ is the
total energy of the $\beta$ particle, $W_0$ is the total energy
available in $m_e c^2$ units, and $p=\sqrt{W^2 -1}$ is the momentum
in $m_e c$ units.
This function weights differently the strength $B(GT)$ depending on the 
excitation energy. As a general rule  $f(Z,Q_{\beta}-E_{ex})$
increases with the energy of the $\beta$-particle and therefore
the strength located at low excitation energies contributes more
significantly to the half-life.

\begin{figure}[ht]
\centering
\includegraphics[width=80mm]{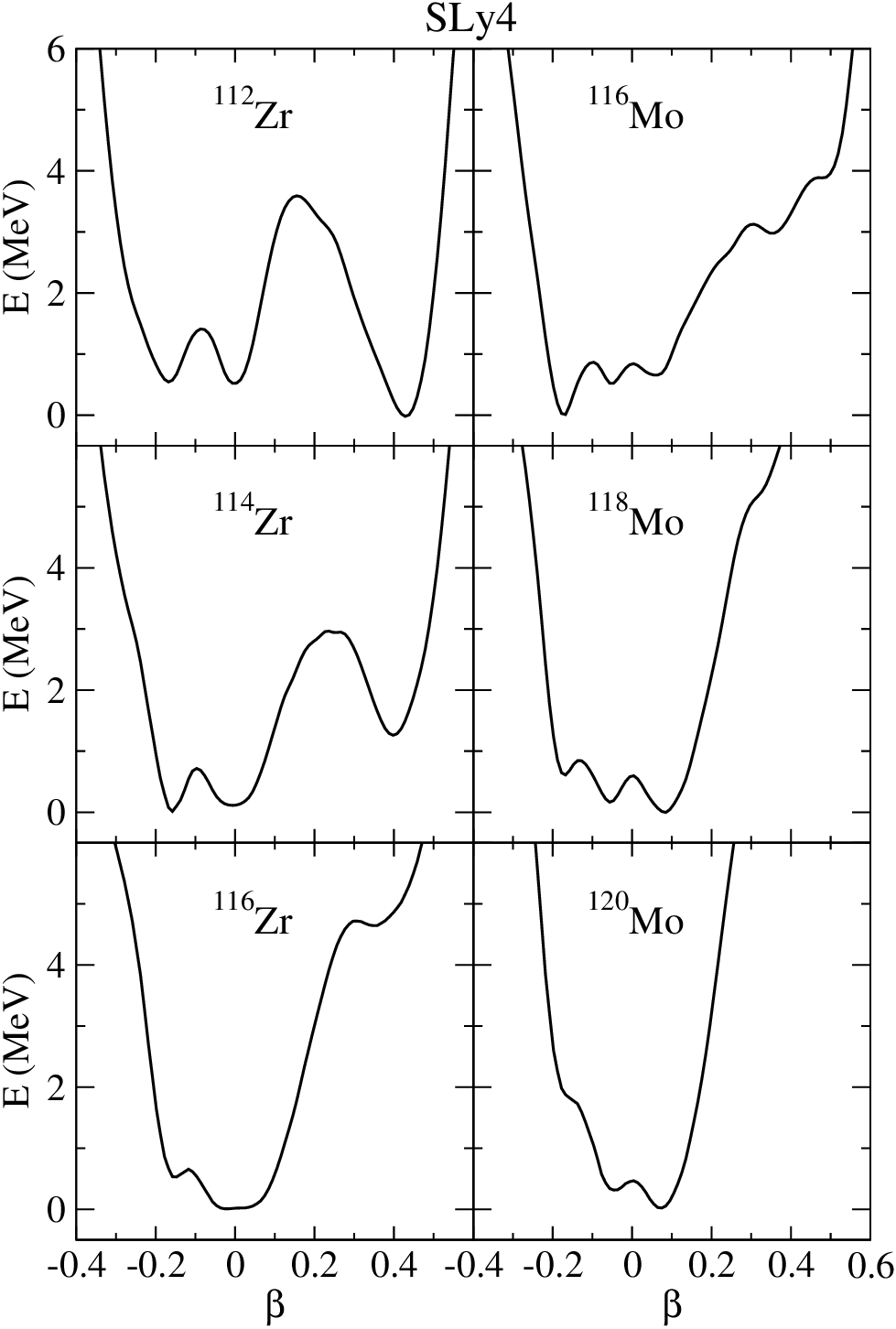}
\caption{Potential energy curves for even-even $^{112,114,116}$Zr and
$^{116,118,120}$Mo isotopes obtained from constrained HF+BCS calculations 
with the Skyrme force SLy4.}
\label{fig_eq}
\end{figure}

\begin{figure}[ht]
\centering
\includegraphics[width=80mm]{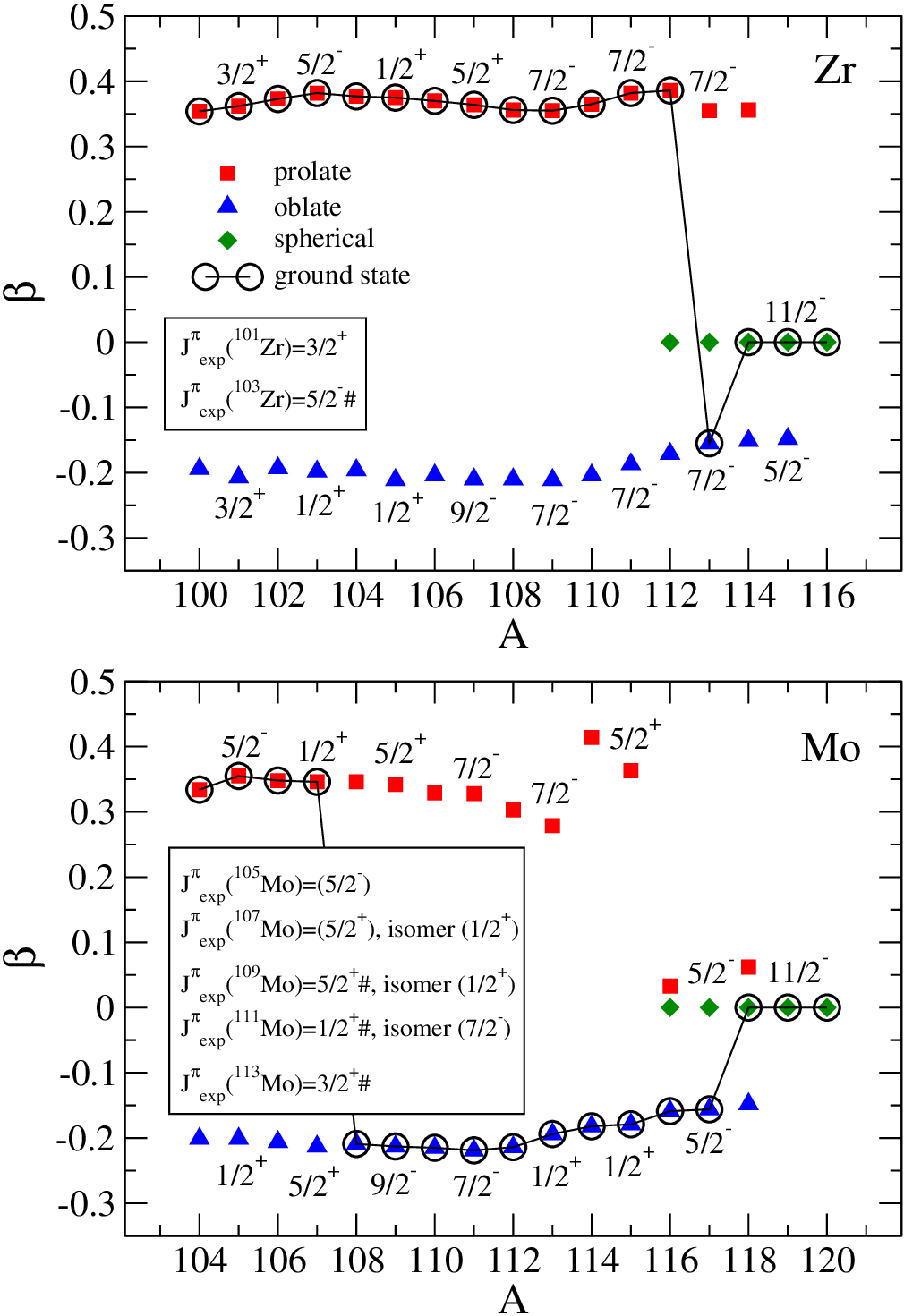}
\caption{(Color online) Isotopic evolution of the quadrupole
deformation parameter $\beta$ of the energy minima obtained from
the Skyrme interaction SLy4 for Zr and Mo isotopes.}
\label{fig_beta}
\end{figure}

\begin{figure}[ht]
\centering
\includegraphics[width=80mm]{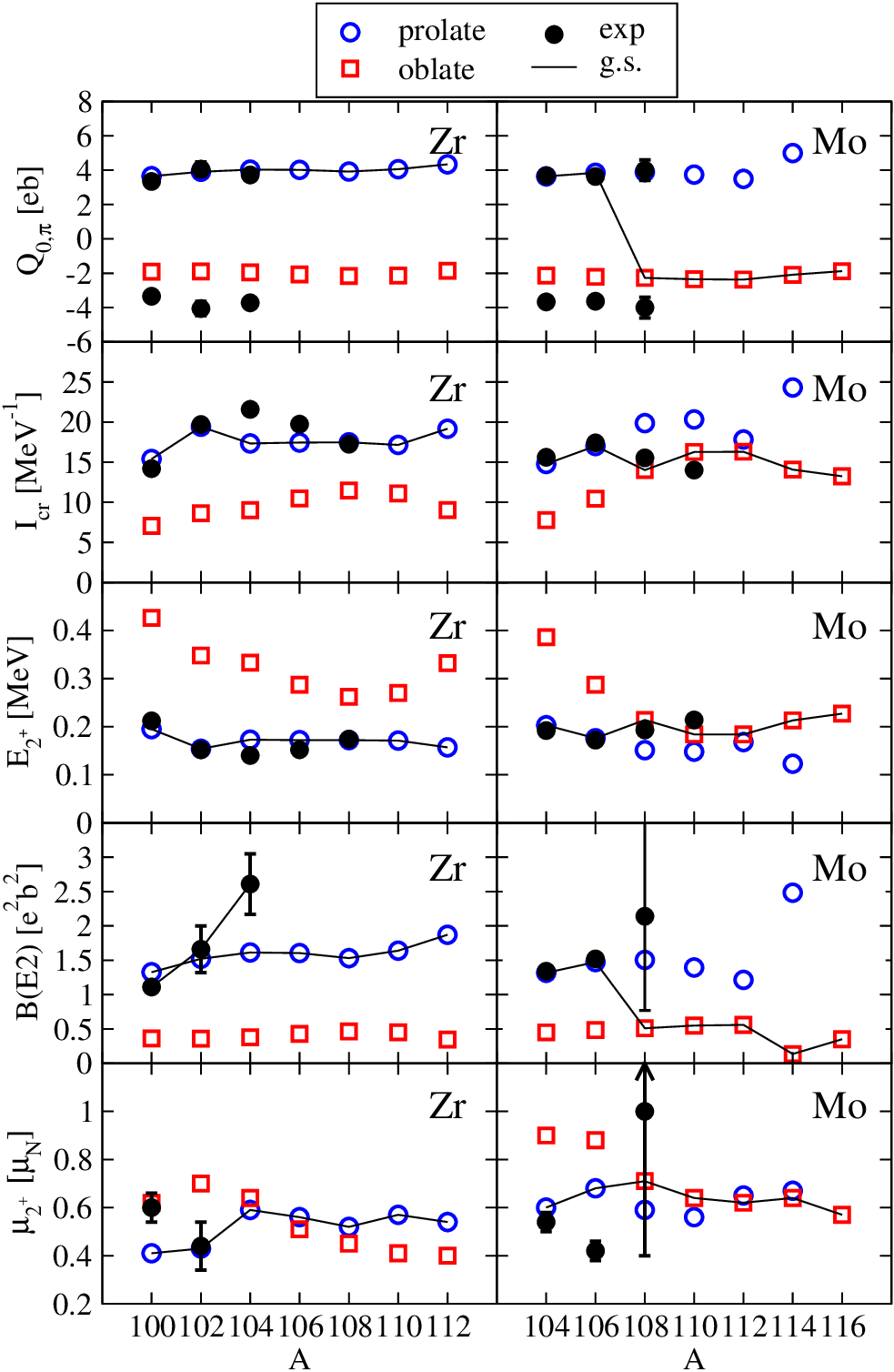}
\caption{(Color online) Isotopic evolution of intrinsic quadrupole moments
($Q_{0,\pi}$), moments of inertia ($\cal I$), excitation energies of $2^+$ states, 
$B(E2)$ transition probabilities, and magnetic dipole moments ($\mu_{2^+}$) in 
even-even Zr and Mo isotopes. Experimental data are from \cite{raman,stone,ensdf}. }
\label{fig_em_even}
\end{figure}

\begin{figure}[ht]
\centering
\includegraphics[width=80mm]{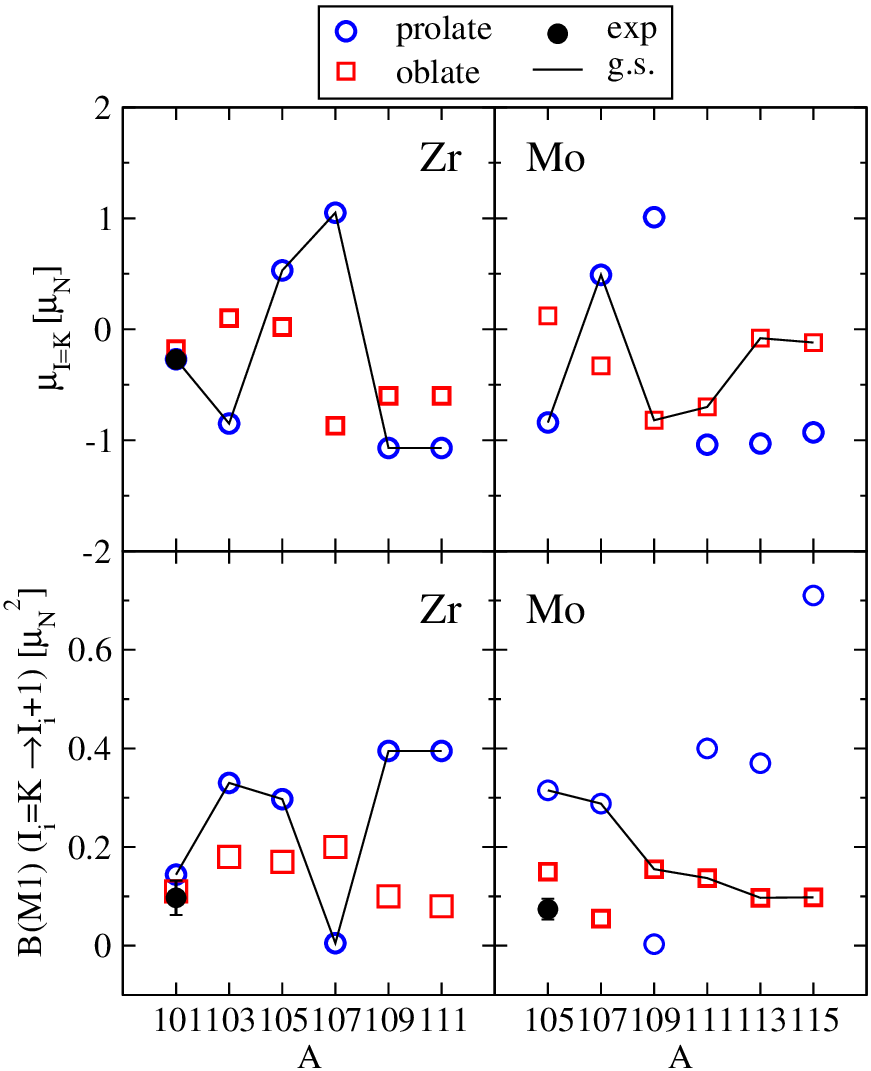}
\caption{(Color online)  Isotopic evolution of  magnetic dipole moments
($\mu_{I=K}$) and  $B(M1)$ transition probabilities in odd-$A$  Zr and Mo 
isotopes. Experimental data are from \cite{stone,ensdf}. }
\label{fig_em_odd}
\end{figure}

The probability for $\beta$-delayed neutron emission is given by

\begin{equation}
P_n = \frac{ {\displaystyle \sum_{S_n < E_{ex} < Q_\beta}f\left( Z,Q_{\beta}-E_{ex}
\right) B(GT,E_{ex}) }}
{{\displaystyle \sum_{0 < E_{ex} < Q_\beta}f\left( Z,Q_{\beta}-E_{ex} \right)
B(GT,E_{ex})}}\, ,
\label{pn}
\end{equation}
where the sums extend to all the excitation energies in the daughter
nuclei in the indicated ranges. $S_n$ is the one-neutron separation
energy in the daughter nucleus. In this expression it is assumed that
all the decays to energies above $S_n$ in the daughter nuclei lead always
to delayed neutron emission and then, $\gamma$-decay from
neutron unbound levels is neglected. According to Eq.~(\ref{pn}), 
$P_{n}$ is mostly sensitive to the strength located at energies around 
$S_n$, thus providing a structure probe complementary to $T_{1/2}$.

\section{Results and discussion}
\label{results}

In this section we start by showing the results obtained for the
potential energy curves in the most unstable isotopes not considered
in Ref. \cite{sarri_pere}. Then, we calculate the energy distribution of 
the GT strength corresponding to the local minima in the potential 
energy curves. After showing the predictions of various models for
the $Q_\beta$ and $S_n$ values for the most unstable isotopes where no 
data are available, we calculate $T_{1/2}$ and $P_{n}$, and discuss the 
results.

\subsection{Potential Energy Curves}

PECs obtained from constrained HF+BCS calculations for even-even isotopes 
$^{100-110}$Zr and $^{104-114}$Mo were discussed in Ref. \cite{sarri_pere}. 
In that reference, it was shown that both Zr and Mo isotopes exhibit two 
well developed oblate and prolate minima, which are separated by barriers 
ranging from 3 MeV up to 5 MeV. In the case of Zr isotopes, the ground 
states are located in the prolate sector at positive values of 
$\beta \approx 0.4$. There are also oblate minima at somewhat higher 
energies located at $\beta \approx -0.2$. In the isotopes $^{108-110}$Zr a 
spherical local minimum is also developed. For Mo isotopes we observed 
practically degenerate oblate and prolate shapes in the light $^{104,106}$Mo 
isotopes, oblate ground states in heavier isotopes with quadrupole 
deformations at $\beta \approx -0.2$ with prolate excited states at energies 
lower than 1 MeV ($\beta \approx 0.4$), and spherical configurations at very 
low energies in heavier isotopes, resulting in an emergent triple 
oblate-spherical-prolate shape coexistence scenario. These results are in 
qualitative agreement with similar ones obtained in this mass region from 
different theoretical approaches including macroscopic-microscopic methods 
based on liquid drop models with shell corrections \cite{skalski97,FRDM}, 
relativistic mean fields \cite{lala2}, as well as nonrelativistic calculations 
with Skyrme \cite{bonche85} and Gogny  \cite{hilaire} interactions.
Thus, a consistent theoretical description emerges, which is supported by
the still scarce experimental information available
\cite{mach,goodin,urban,campbell,charlwood}.

In this work, we complete the picture with the inclusion of the PECs in 
$^{112,114,116}$Zr and $^{116,118,120}$Mo isotopes. We show in Fig. \ref{fig_eq} 
the energies relative to that of the ground state plotted as a function of 
the quadrupole deformation $\beta$. The general trend observed is the 
gradual disappearance of both prolate and oblate minima collapsing into a 
spherical solution in the heavier isotopes as the magic number $N=82$ is 
approached.

To further illustrate the role of deformation in the isotopic evolution,
we show in Fig. \ref{fig_beta} the quadrupole deformation $\beta$ of the 
various energy minima as a function of the mass number $A$, for Zr and Mo 
isotopic chains. The deformation corresponding to the ground state for 
each isotope is encircled. We can see that in most cases we have two 
minima, in the prolate and oblate sectors, which are very close in energy.
It is also interesting to compare the spin and parity ($J^\pi$) of the 
different shapes with the experimental assignments. We can see in 
Fig. \ref{fig_beta} those values for the odd-$A$ isotopes.
In the case of Zr isotopes, the experimental $J^\pi$ are $3/2^+$ and
$5/2^- \#$ for $^{101}$Zr and $^{103}$Zr, respectively. 
The symbol $\#$ indicates here and in what follows that the values are 
estimated from trends in neighboring nuclides. They correspond 
nicely with the spin and parities obtained with SLy4 for the lowest 
one quasiparticle prolate states.  In the case of Mo isotopes, $^{105}$Mo is
a $(5/2^- )$ state well described by the prolate SLy4 ground state. 
$^{107}$Mo is a $(5/2^+)$ state with a $(1/2^+)$ excited state at 65 keV.
They appear as prolate ($1/2^+$) and oblate ($5/2^+$) states in the 
calculations at very close energies.
In heavier isotopes some discrepancies are found between the spin-parity
of the calculated ground states and the experimental assignments 
obtained from systematics. Nevertheless, the spin and parity of $^{109}$Mo 
($5/2^+ \#$) corresponds to the prolate shape in the calculations, 
whereas in $^{111}$Mo the spin and parity of the isomer state  ($7/2^-$) 
agrees with the oblate configuration.

\begin{figure}[ht]
\centering
\includegraphics[width=75mm]{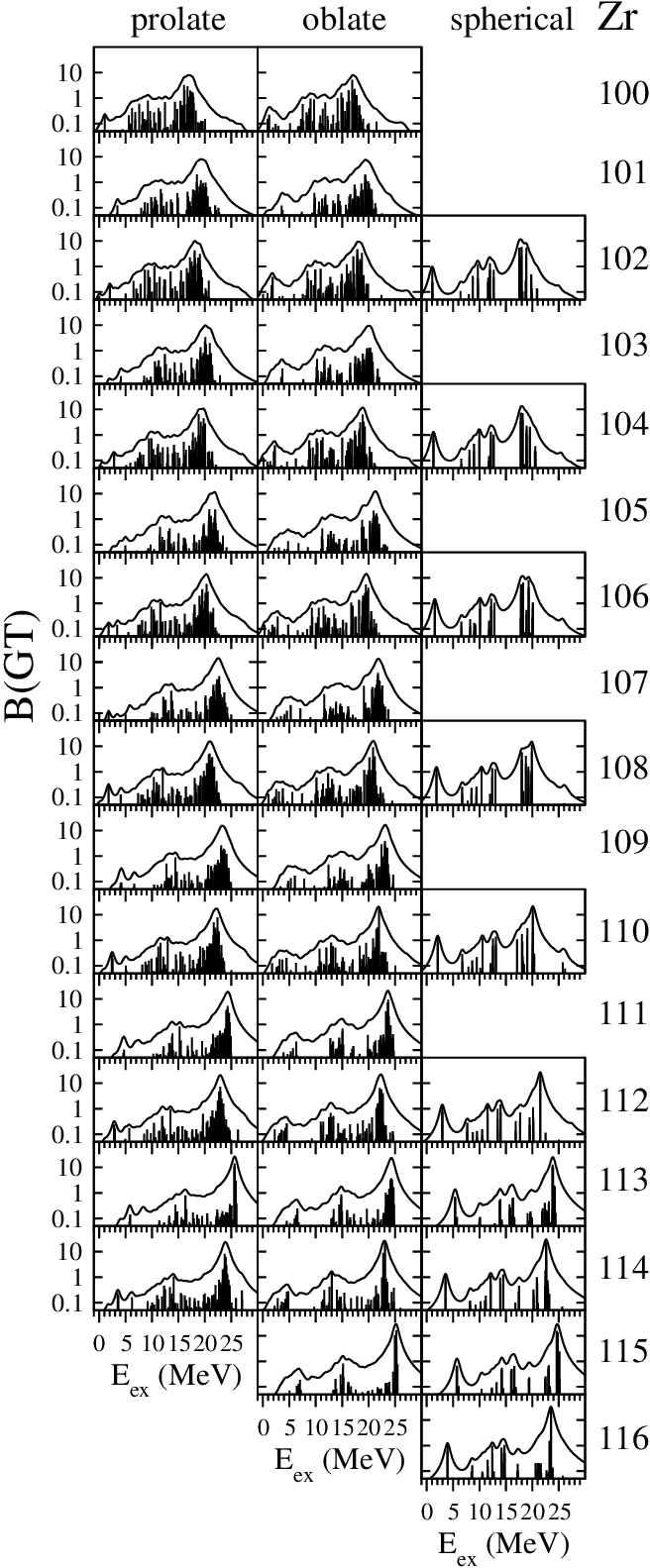}
\caption{(Color online) QRPA-SLy4 Gamow-Teller strength distributions 
for Zr isotopes as a function of the excitation energy in the daughter
nucleus. The calculations correspond to the various equilibrium
configurations found in the PECs.}
\label{fig_bgt_zr}
\end{figure}

\begin{figure}[ht]
\centering
\includegraphics[width=75mm]{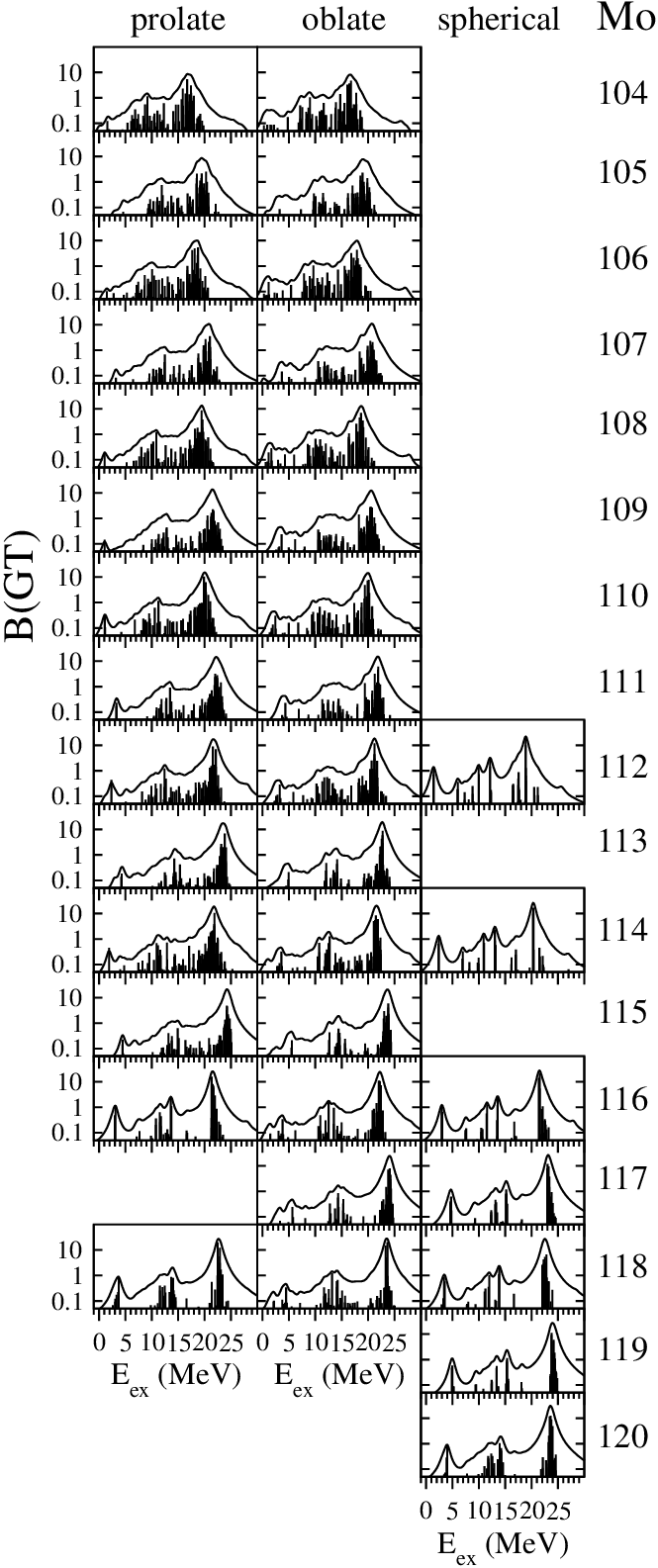}
\caption{(Color online) Same as in  Fig. \ref{fig_bgt_zr}, but for Mo 
isotopes.}
\label{fig_bgt_mo}
\end{figure}

\subsection{Electromagnetic properties}

Although the main focus of this work is the decay properties of neutron-rich
Zr and Mo isotopes, it is also worth discussing their electromagnetic 
properties within the rotational formalism of Bohr and Mottelson \cite{bm}. 
Then, we discuss the isotopic 
structural evolution in these chains by studying electric quadrupole 
moments  $Q_{0,\pi}$, moments of inertia $\cal I$, excitation energies of 
the $2^+$ excited states $E_{2^+}$, $B(E2)$ reduced transition probabilities 
and magnetic dipole moments $\mu_I$, in the case of even-even isotopes. 
In the case of odd-$A$ isotopes we concentrate the study on  magnetic dipole 
moments and $B(M1)$ reduced transition probabilities.

In Fig. \ref{fig_em_even} we present these results for the even Zr and Mo
isotopes. We show results for both prolate and oblate shapes because 
they have very close energies. The ground state shapes predicted in the 
calculation are connected with solid lines. They correspond to prolate 
shapes in the Zr isotopes and to prolate shapes in $^{104,106}$Mo
isotopes and oblate shapes in  $^{108-116}$Mo isotopes.
The intrinsic electric quadrupole moments are calculated microscopically 
from the SLy4 Skyrme interaction used in this work.
The moments of inertia $\cal I$ are calculated in a cranking approach as
described in \cite{moya} and the rotational energies of the states $I$ 
of the ground state band $K=0$ are given in terms of  $\cal I$ by

\begin{equation}
E^I_{rot}=\frac{1}{2\cal I} [ I(I+1) ] \, .
\end{equation}
The reduced transition probability corresponding to a $E2$ transition within a 
band $K$ is given by

\begin{equation}
B(E2,I_iK \rightarrow I_fK)=\frac{5}{16\pi}\left< I_i 2 K 0| I_f K\right> ^2 
e^2 Q_{0,\pi}^2\, .
\end{equation}
For the transition from the ground state $0^+$ to the $2^+$ rotational state 
this expression reduces to

\begin{equation}
B(E2,0^+ \rightarrow 2^+)=\frac{5}{16\pi} e^2 Q_{0,\pi}^2\, .
\end{equation}
The magnetic dipole moment of a rotational state is given by

\begin{eqnarray}
\mu_I&=&g_R I + \left( g_K-g_R \right) \frac{K^2}{I+1} \{ 1+  \nonumber \\
&&  \delta_{K,1/2} (-1)^{I+1/2}(2I+1) b \}\, ,
\end{eqnarray}
in terms of the rotational ($g_R$) and single-particle ($g_K$) gyromagnetic 
ratios, and the magnetic decoupling parameter $b$ \cite{bm}.
For $K=0$ even-even isotopes  $\mu_I$ is simply given by $\mu_I=g_R I $.

In the case of odd-$A$ isotopes we calculate magnetic dipole moments and 
$M1$ reduced transition probabilities connecting states within a rotational 
band,

\begin{eqnarray}
&&B(M1, I_i\ K \rightarrow I_f\ K)=\frac{3}{4\pi}\left< I_i1K0|I_fK\right>^2 \nonumber \\
&& \left(g_K-g_R\right) ^2 K^2 \left\{ 1+\delta_{K,1/2}(-1)^{I_>+1/2}b\right\}^2 \mu_N^2
\end{eqnarray}
In Fig. \ref{fig_em_odd} we show the results corresponding to $\mu_{I=K}$ and 
$B(M1)$ for transition probabilities connecting the ground states $I_i=K$ with
the first excited states in the rotational band $I_f=I_i+1$.

Data are taken from Refs. \cite{raman,stone,ensdf}. The intrinsic quadrupole
moments have been plotted with the two possible signs because they
are extracted from $B(E2)$ values \cite{raman}.
The agreement with
the still scarce experimental information is very reasonable, favoring the
description of the lighter Zr and Mo isotopes as prolate in agreement with
the calculations. Only the data on  $Q_{0,\pi}$ and $B(E2)$ in $^{108}$Mo 
seem to be at variance with the calculation that predicts an oblate shape 
for the ground state, whereas the data seem to favor a prolate shape,
thus shifting the expected transition from prolate to oblate to a heavier 
isotope.

\subsection{Gamow-Teller strength distributions}

In the next figures, we show the results obtained for the energy
distributions of the GT strength corresponding to the 
oblate-prolate-spherical equilibrium shapes for which we obtained
relevant minima in the PECs. The results are obtained from QRPA 
with the force SLy4 with pairing correlations and with residual 
interactions with the parameters written in Sec. \ref{sec2}.  The GT 
strength in $(g_A^2/4\pi)$ units, is plotted versus the excitation 
energy of the daughter nucleus and a quenching factor 0.77 has been 
included.

\begin{figure}[ht]
\centering
\includegraphics[width=80mm]{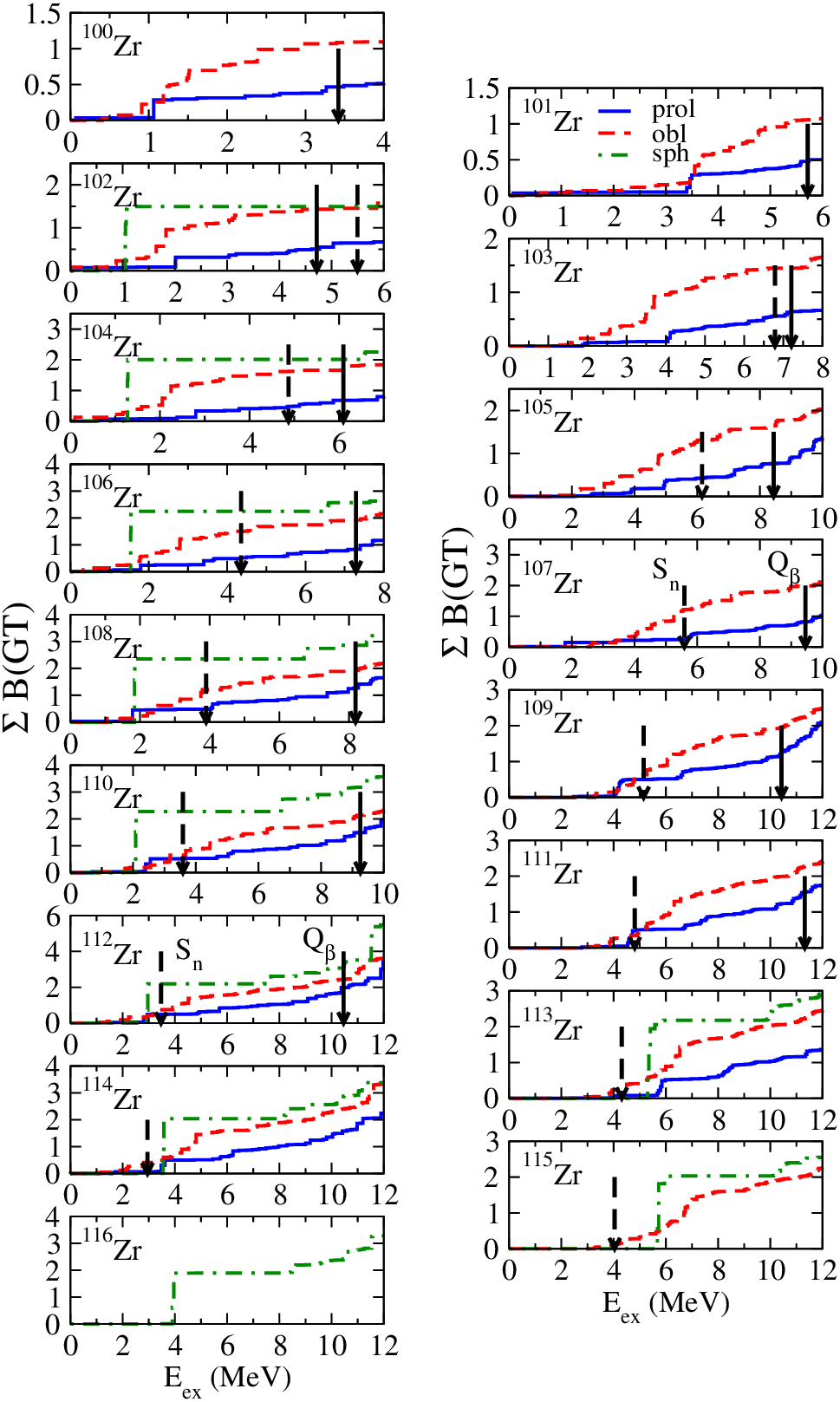}
\caption{(Color online) QRPA-SLy4 accumulated GT strengths in Zr isotopes 
calculated for the various equilibrium shapes. 
$Q_\beta$ and $S_n$ energies are shown by solid and dashed vertical 
arrows, respectively.}
\label{fig_bgt_zr_qbeta}
\end{figure}

\begin{figure}[ht]
\centering
\includegraphics[width=80mm]{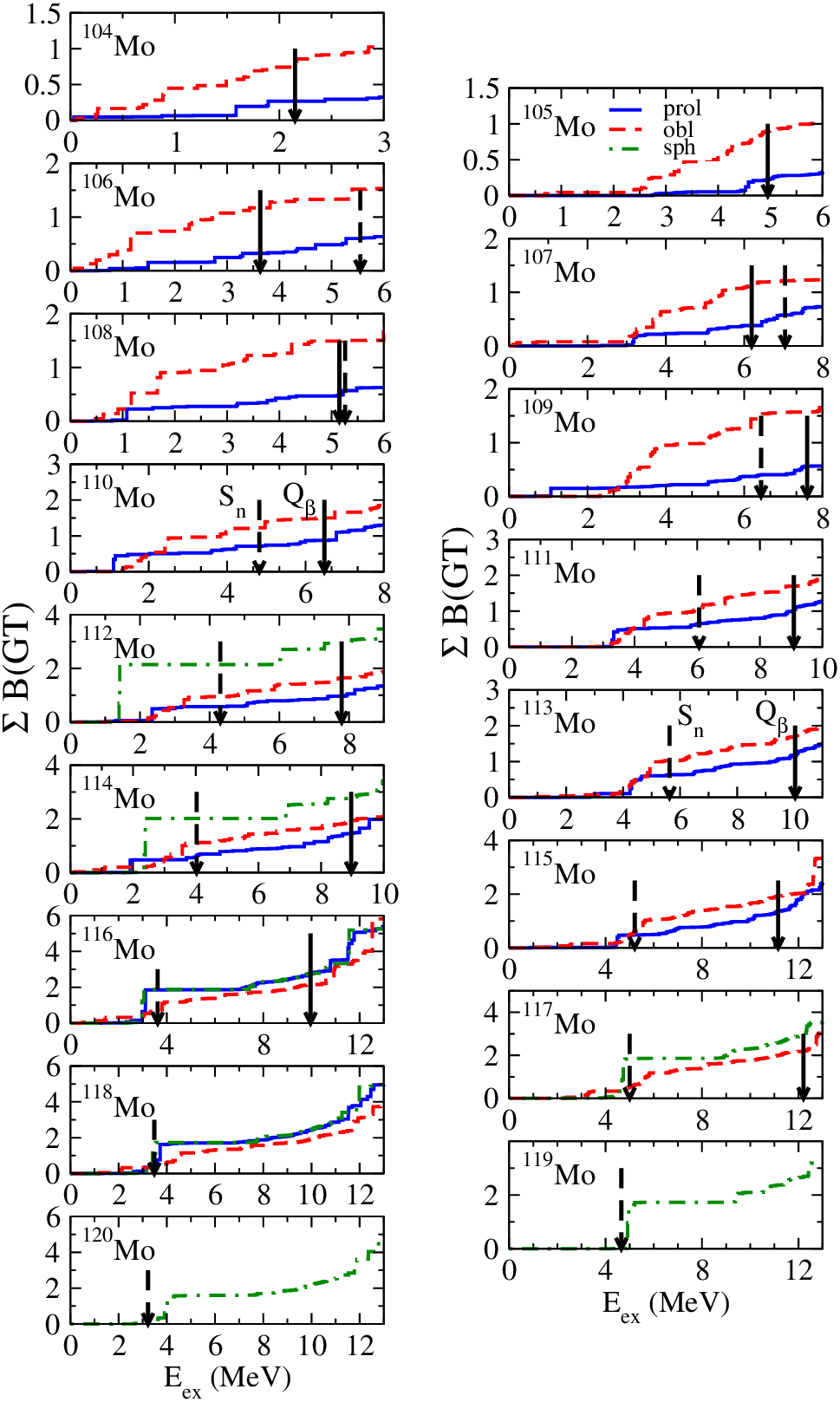}
\caption{(Color online) Same as in Fig. \ref{fig_bgt_zr_qbeta}, but for Mo 
isotopes.}
\label{fig_bgt_mo_qbeta}
\end{figure}

Figs. \ref{fig_bgt_zr} and \ref{fig_bgt_mo} contain the results for Zr
and Mo isotopes, respectively. We show the energy distributions of the
individual GT strengths together with continuous distributions obtained
by folding the strength with 1 MeV width Breit-Wigner functions.
The main characteristic of these distributions is the existence of a GT
resonance located at increasing excitation energy as the number of neutrons
$N$ increases. The total GT strength also increases with $N$, as it is
expected to fulfill the Ikeda sum rule. It is worth noticing that both
oblate and prolate shapes produce quite similar GT strength distributions
on a global scale. Nevertheless, the small differences among the various 
shapes at the low energy tails (below the $Q_\beta$) of the GT strength 
distributions that can be appreciated because of the logarithmic scale, 
lead to sizable effects in the $\beta$-decay half-lives. In the next 
figures, Fig. \ref{fig_bgt_zr_qbeta} for Zr isotopes and 
Fig. \ref{fig_bgt_mo_qbeta} for Mo isotopes, we can see the accumulated 
GT strength in the energy region below the corresponding $Q_\beta$ energy 
of each isotope, which is the relevant energy range for the calculation 
of the half-lives. The vertical solid (dashed) arrows show the $Q_\beta$ 
($S_n$) energies, taken from experiment \cite{audi12}.

\begin{figure}[ht]
\centering
\includegraphics[width=80mm]{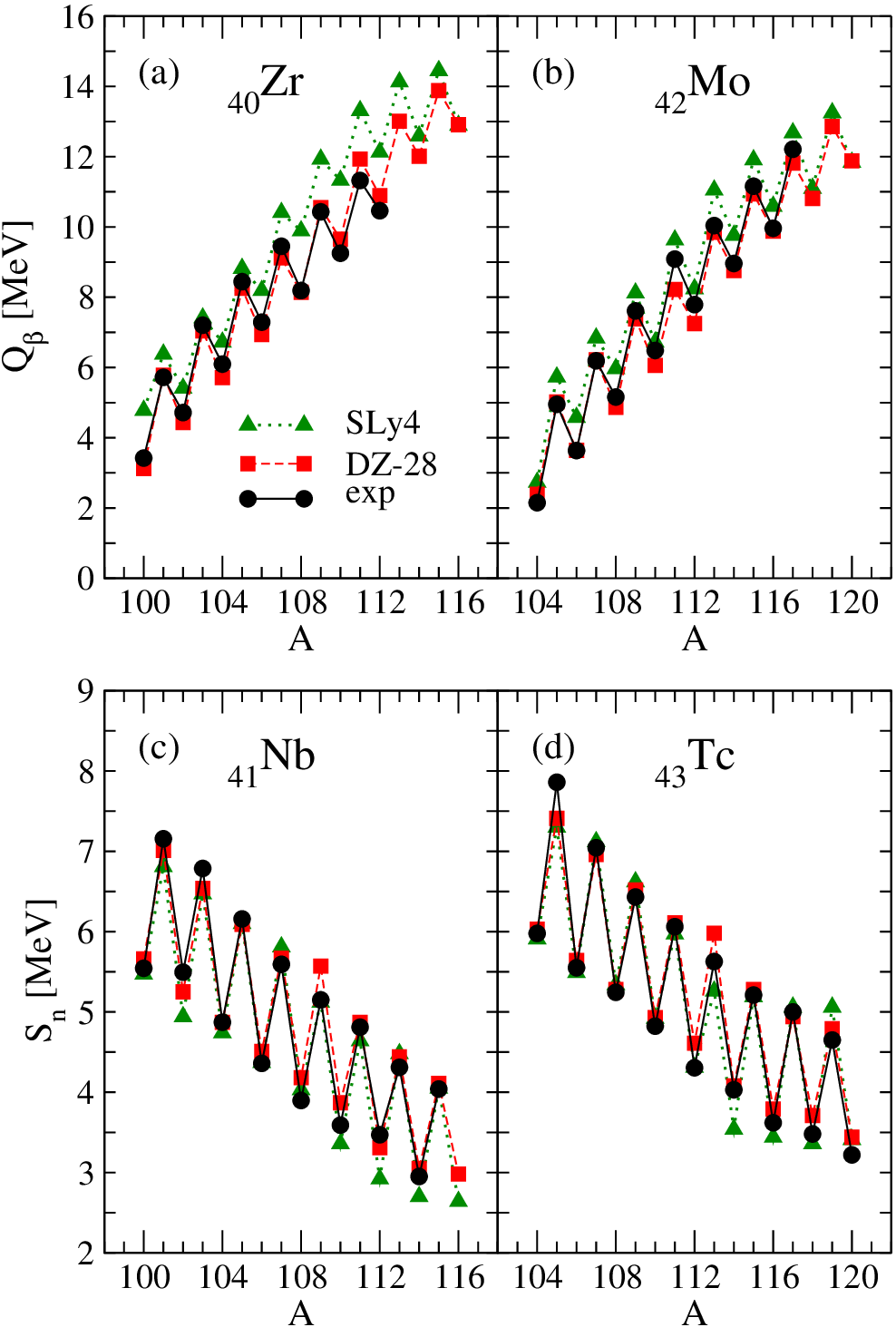}
\caption{(Color online) Experimental $Q_\beta$ and $S_n$ energies compared to the
predictions of various mass models.}
\label{fig_qbeta_sn_red}
\end{figure}

In these figures one can appreciate the sensitivity of these distributions 
to deformation and how measurements of the GT strength distribution from 
$\beta$-decay can be a powerful tool to get information about this deformation, as 
it was carried out in Refs. \cite{exp_poirier,exp_nacher}. The accumulated 
strength from the oblate shapes is in most cases larger than the 
corresponding prolate profiles. The spherical distributions have distinct 
characteristics showing always as a strong peak at relative low energies. 
The profiles from different shapes could be easily distinguished 
experimentally from each other. This is specially true in the case of the 
lighter isotopes $^{100-104}$Zr and  $^{104-108}$Mo, where the differences are 
enhanced. These isotopes are in principle easier to measure since they are 
closer to stability.

Experimental information on GT strength distributions in these isotopes
is only available in the energy range below 1 MeV for the isotopes
$^{106,108}$Mo \cite{jokinen}, $^{110}$Mo \cite{wang04}, and
$^{100,102,104}$Zr \cite{rinta07}. 
Unfortunately, the energy region is still very narrow and represents
only a small fraction of the GT strength relevant for the half-life
determination. Clearly, more experimental information is needed
to get insight into the nuclear structure of these isotopes. 

Knowledge of the energy distribution of the GT strength is of paramount 
importance to understand and constrain the underlying nuclear structure 
responsible for the nuclear GT response. Half-lives are integral quantities
of this strength weighted with appropriate phase factors (see Eq. 
\ref{t12}), but reproducing theoretically the half-lives does not warrant 
the correct description of the GT strength distribution. The latter are 
indeed needed to determine the decay rates in astrophysical scenarios. 
This is because the phase factors are sensitive functions of the electron 
distribution in the medium that can block the available space for the 
$\beta$-particle \cite{langanke}. 
Thus, the stellar phase factors are different from those 
in the laboratory under terrestrial conditions and so are the half-lives. 
Therefore, to describe properly the half-lives under extreme conditions of 
density and the temperature one needs to account not only for genuine 
thermal effects due to the population of excited state in the decaying 
nuclei, but also for a reliable description of the GT strength 
distributions \cite{sarri09_1,sarri09_2}.

It is also worth noting that the GT strength distribution in odd-$A$ 
isotopes is typically displaced to higher energies (about 2-3 MeV) 
with respect to the even-even case. This can be clearly seen in the 
position of the GT resonance in Figs. \ref{fig_bgt_zr}-\ref{fig_bgt_mo} 
or in the lower excitation energy at which we found significant strength 
in Figs. \ref{fig_bgt_zr_qbeta}-\ref{fig_bgt_mo_qbeta}. The shift 
corresponds roughly to the breaking of a neutron pair and therefore it 
amounts to about twice the neutron pairing gap as it can be seen in Eqs.
(\ref{eexeven})-(\ref{eex3qp}). Similarly, one observes that this energy 
shift is the step in the odd-even staggering of $Q_\beta$ ans $S_n$ in
Fig. \ref{fig_qbeta_sn_red}. Taking into account both effects (the 
displacement of the strength and the increase of $Q_\beta$ in odd-$A$ 
nuclei), the final half-lives cancel them to a large extent, giving 
rise to a smooth behavior in $A$, as we shall see in the next subsection.

\begin{figure}[ht]
\centering
\includegraphics[width=80mm]{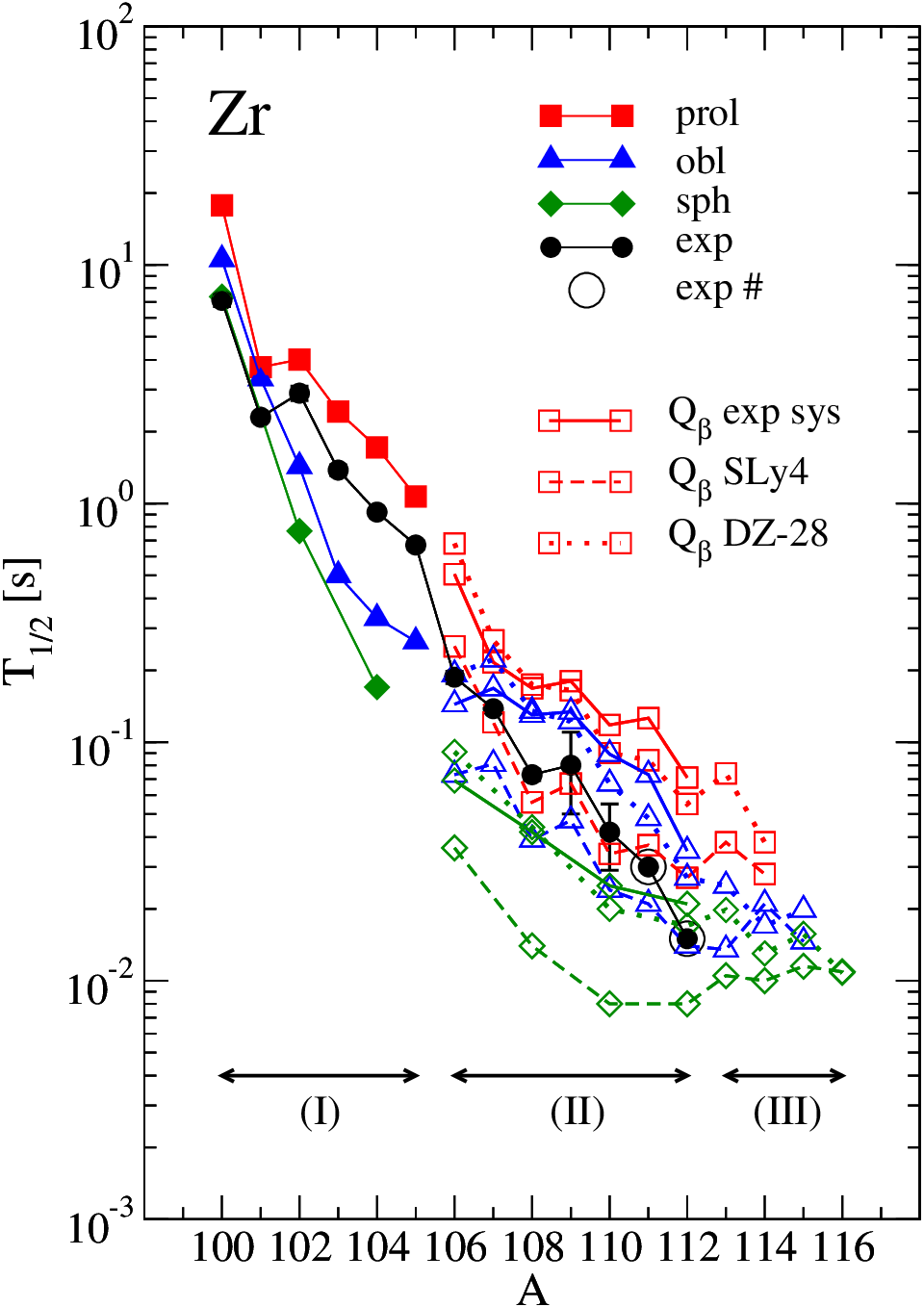}
\caption{(Color online) Measured  $\beta$-decay half-lives for Zr isotopes 
compared to theoretical QRPA-SLy4 results calculated from different shape 
configurations. Solid symbols (I) correspond to half-lives obtained with 
experimental $Q_\beta$ values. Open symbols correspond to results obtained 
with experimental (from systematics), SLy4, and DZ-28 $Q_\beta$ values (II).
In region (III) only results with theoretical $Q_\beta$ values are plotted.}
\label{fig_t_zr_1}
\end{figure}

\begin{figure}[ht]
\centering
\includegraphics[width=80mm]{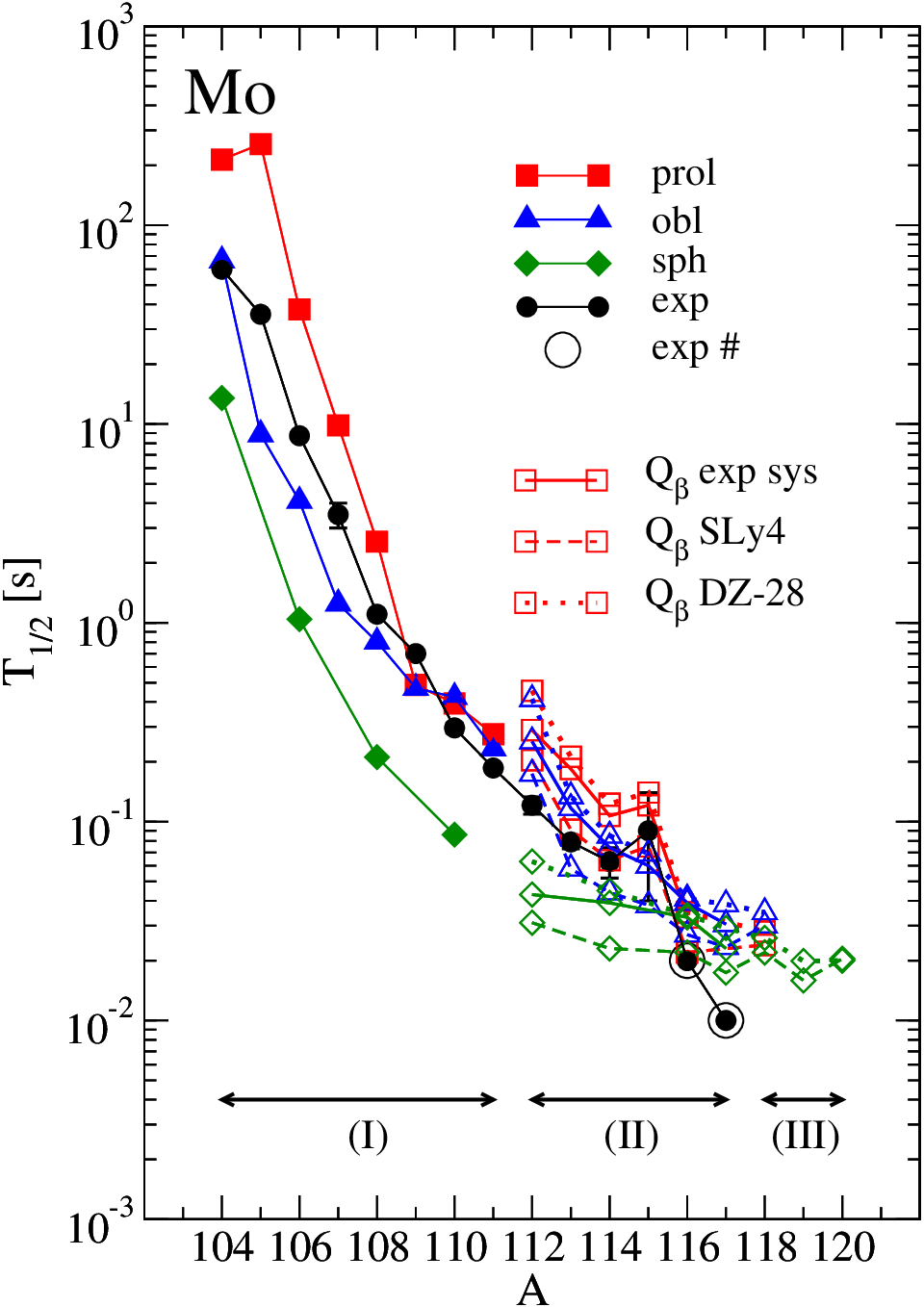}
\caption{(Color online) Same as in Fig. \ref{fig_t_zr_1}, but for
Mo isotopes.}
\label{fig_t_mo_1}
\end{figure}

\begin{figure}[ht]
\centering
\includegraphics[width=80mm]{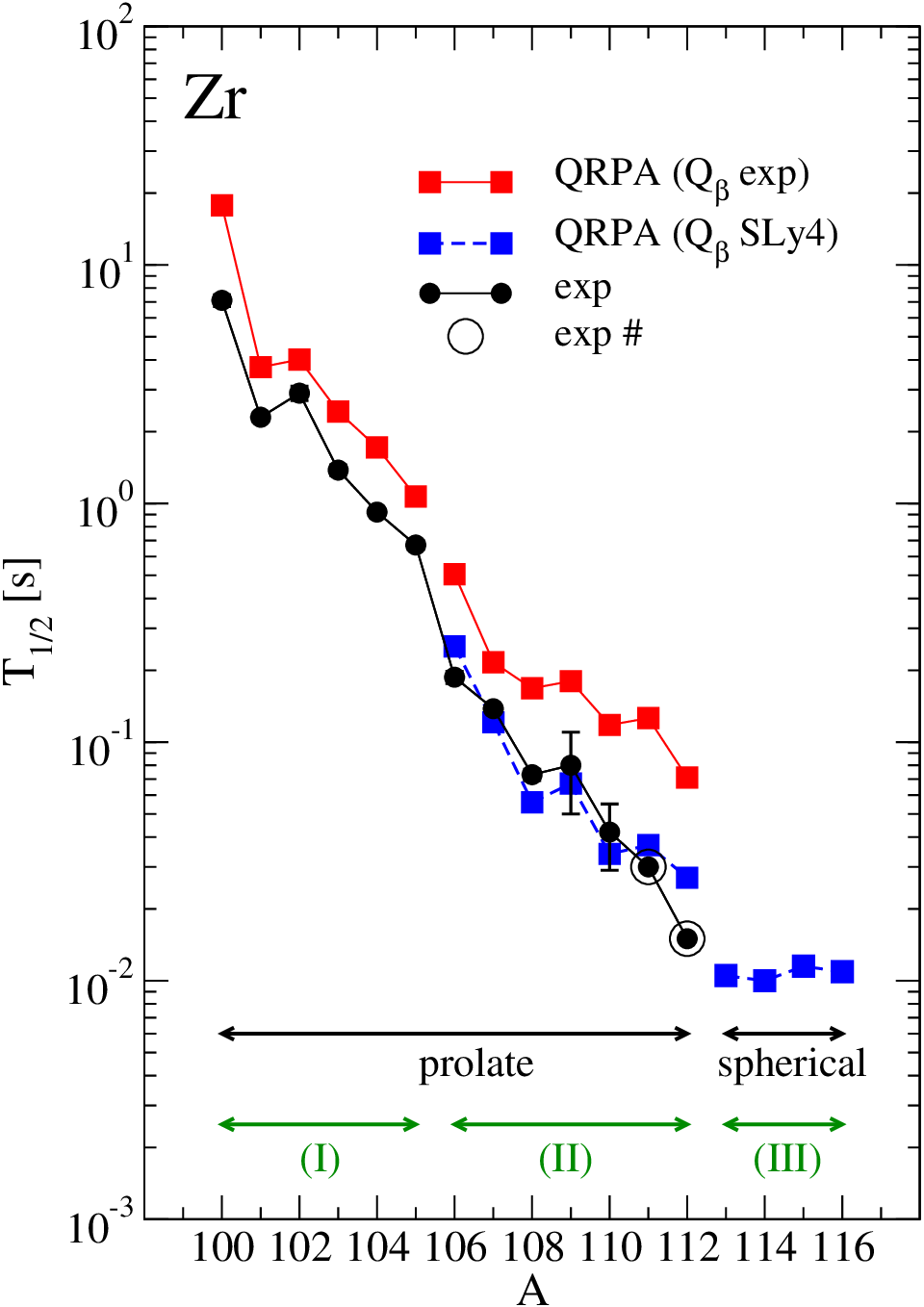}
\caption{(Color online)  Measured  $\beta$-decay half-lives for Zr isotopes 
compared to theoretical QRPA-SLy4 results obtained with the ground state shapes.
The meaning of the three regions (I,II,III) is the same as in Fig. \ref{fig_t_zr_1}}.
\label{fig_t_zr_2}
\end{figure}

\begin{figure}[ht]
\centering
\includegraphics[width=80mm]{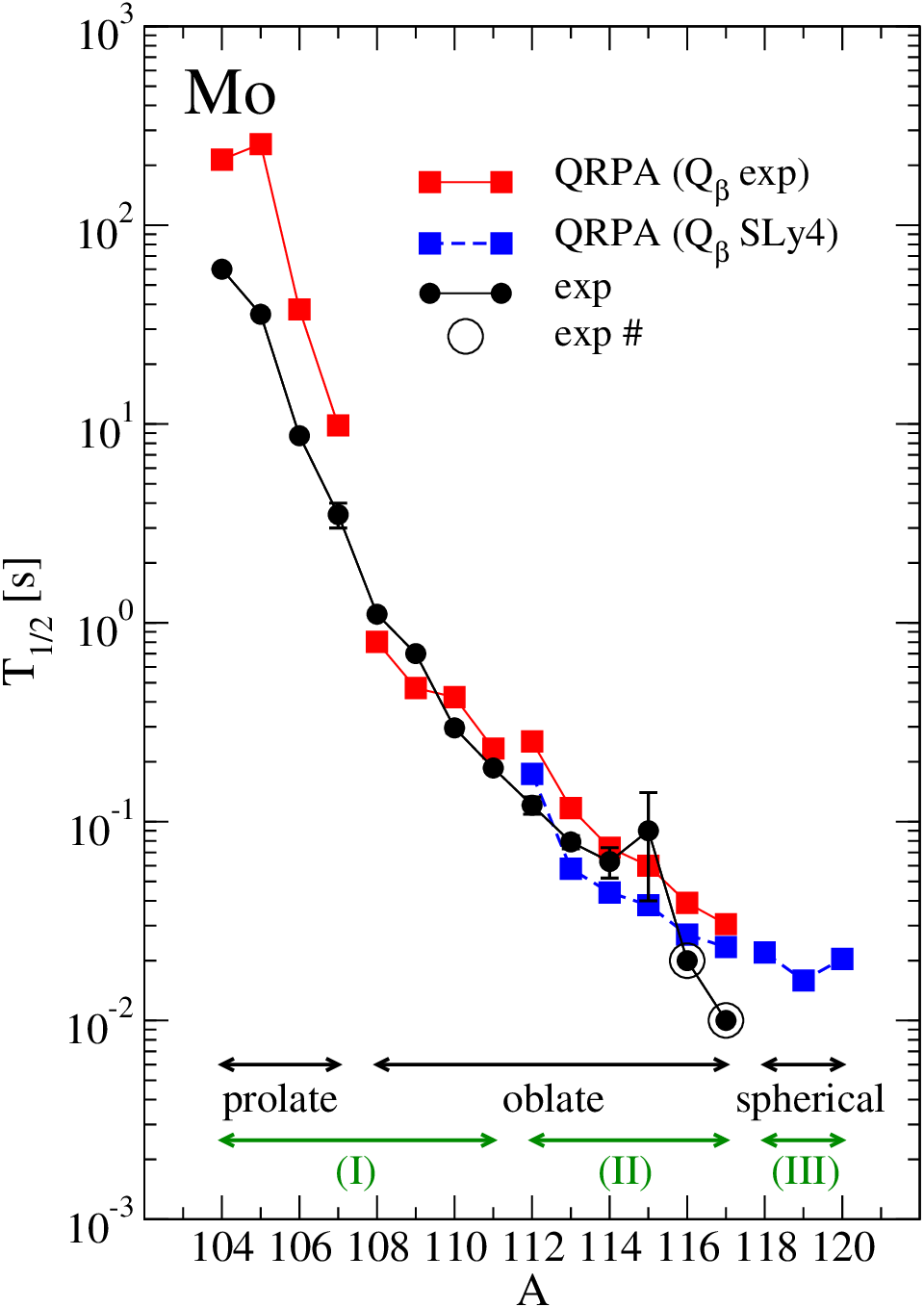}
\caption{(Color online) Same as in Fig. \ref{fig_t_zr_2}, but for Mo isotopes.}
\label{fig_t_mo_2}
\end{figure}

\begin{figure}[ht]
\centering
\includegraphics[width=80mm]{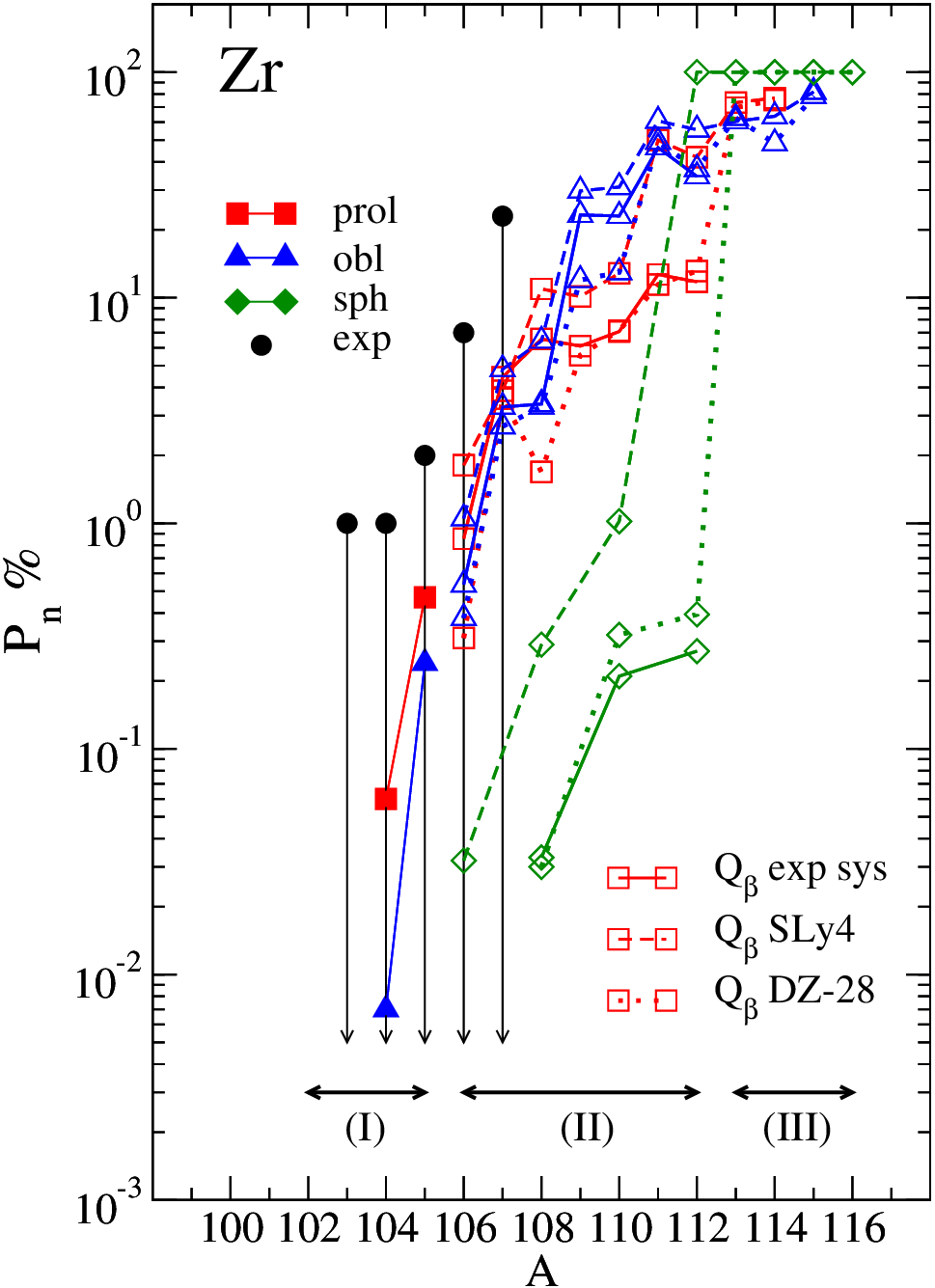}
\caption{(Color online) Same as in Fig. \ref{fig_t_zr_1}, but for 
percentage $P_n$ values.}
\label{fig_pn_zr_1}
\end{figure}

\begin{figure}[ht]
\centering
\includegraphics[width=80mm]{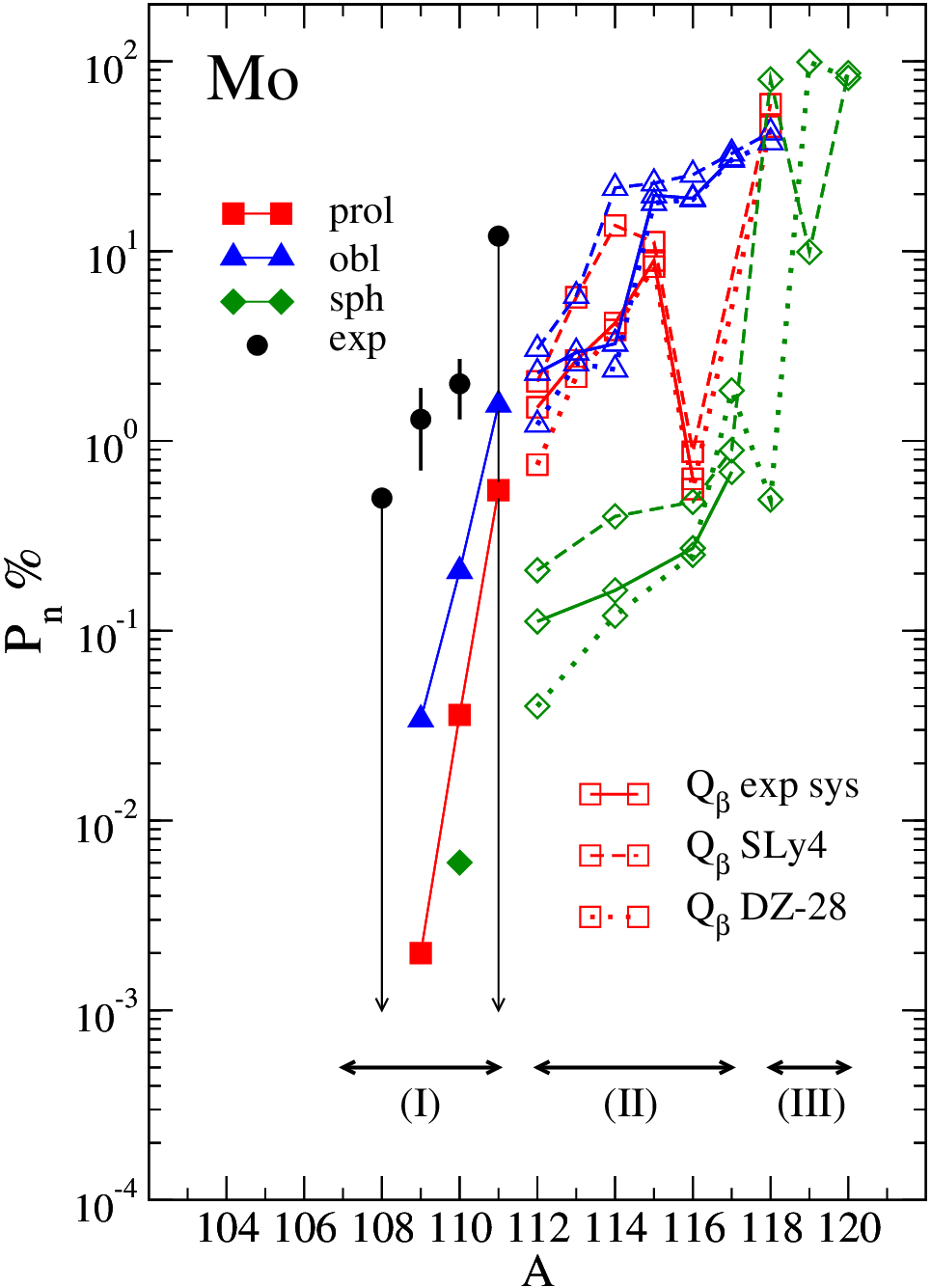}
\caption{(Color online) Same as in Fig. \ref{fig_pn_mo_1}, but for
percentage $P_n$ values.}
\label{fig_pn_mo_1}
\end{figure}

\begin{figure}[ht]
\centering
\includegraphics[width=80mm]{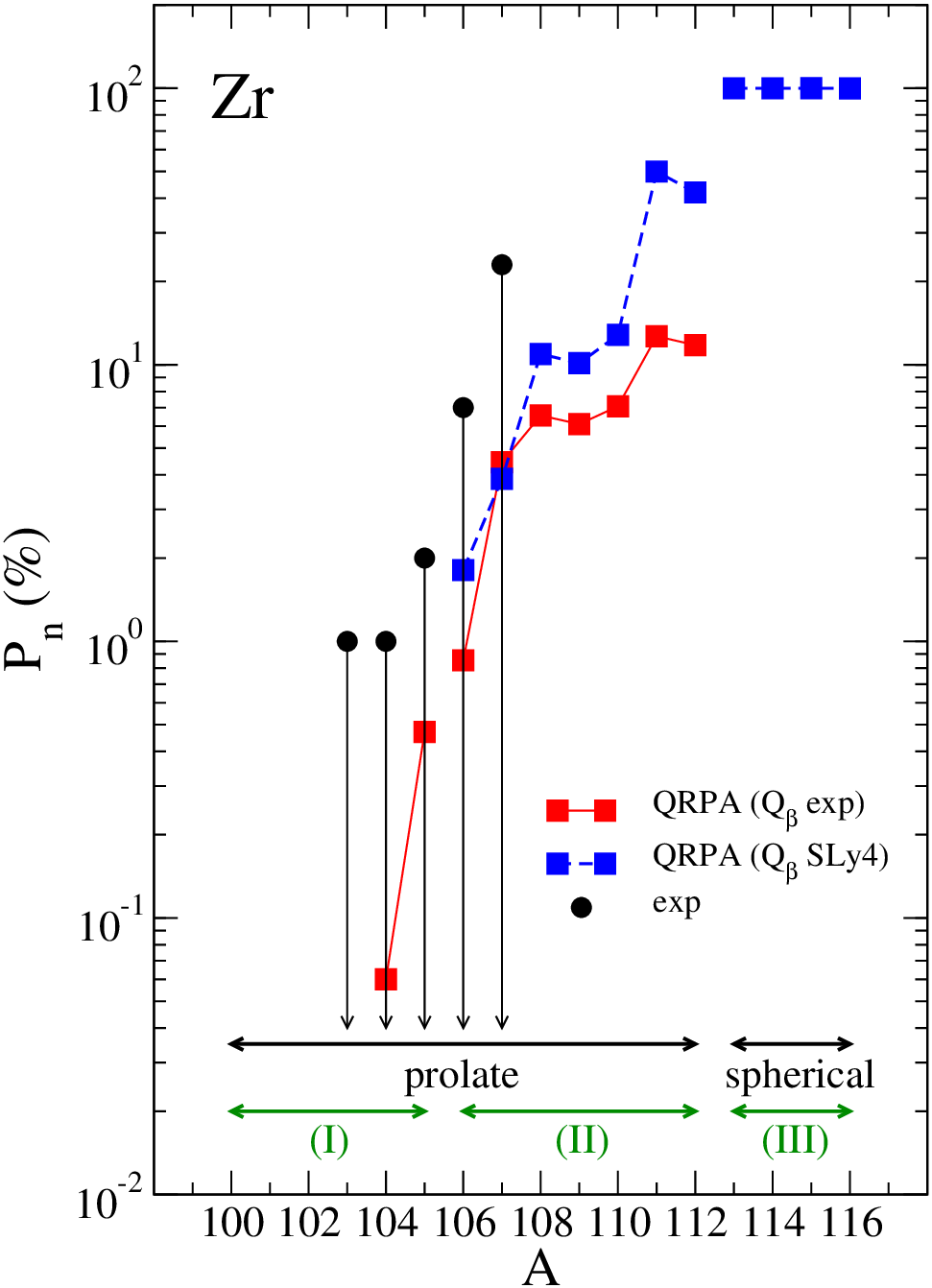}
\caption{(Color online) Same as in Fig. \ref{fig_t_zr_2}, but for $P_n$ values.
}
\label{fig_pn_zr_2}
\end{figure}

\begin{figure}[ht]
\centering
\includegraphics[width=80mm]{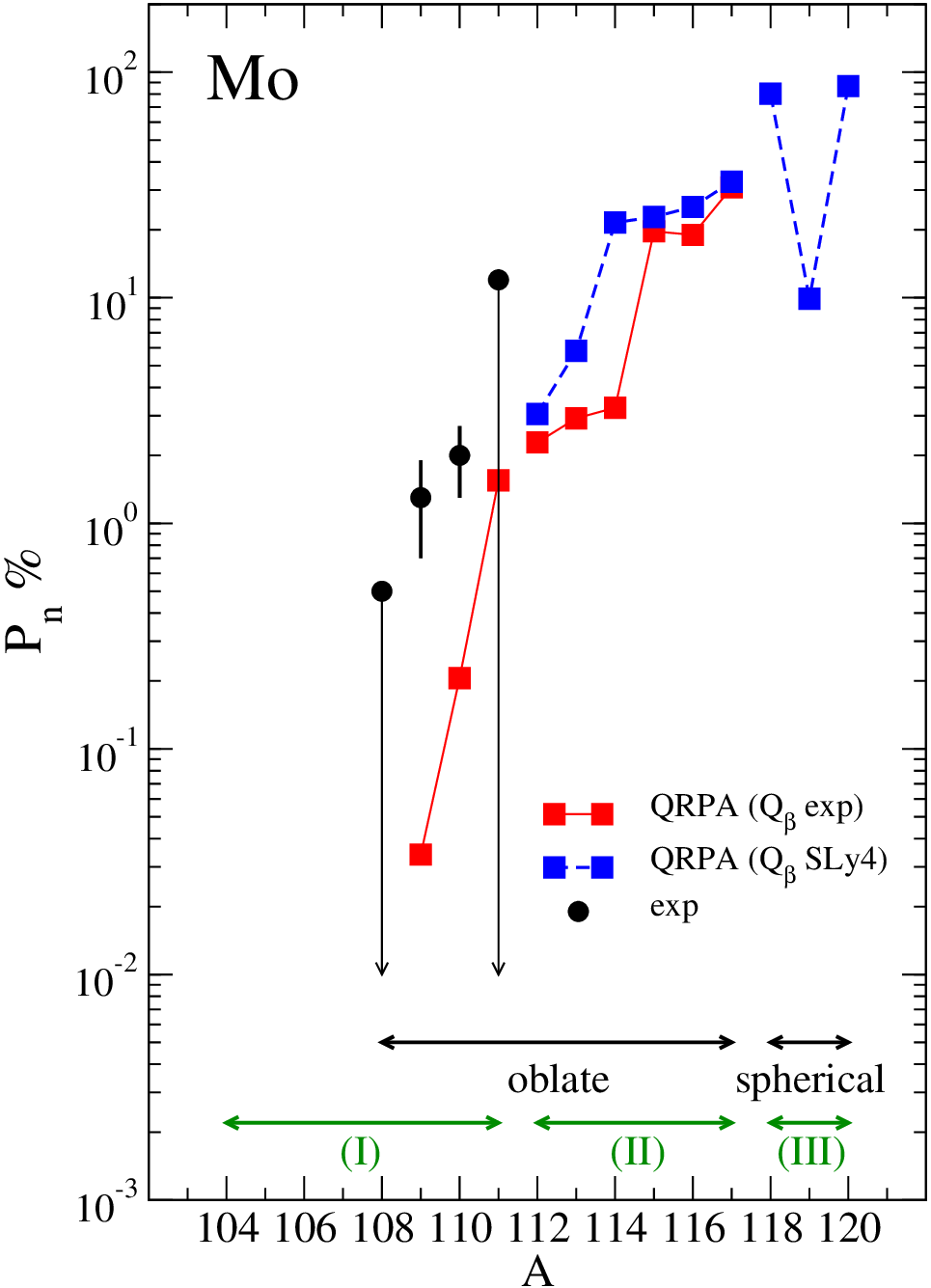}
\caption{(Color online) Same as in Fig. \ref{fig_t_mo_2}, but for
$P_n$ values.}
\label{fig_pn_mo_2}
\end{figure}

\subsection{Half-lives and $\beta$-delayed neutron-emission probabilities}

The calculation of the half-lives in Eq. (\ref{t12}) involves knowledge
of the GT strength distribution and of the $Q_\beta$ values. The calculation 
of the probability for $\beta$-delayed neutron emission $P_n$ in Eq. (\ref{pn})
involves also knowledge of the $S_n$ energies. We use experimental values 
for $Q_\beta$ and $S_n$, which are taken from Ref. \cite{audi12}, when available.
But in those cases where experimental masses are not measured, one has to
rely on theoretical predictions for them.
There are a large number of mass formulas in the market obtained from
different approaches. The study in Ref. \cite{sarri_pere} showed us 
that among them, the Duflo and Zuker (DZ-28) mass model \cite{DufloZuker} 
and the masses calculated from SLy4 with a zero-range pairing force and 
Lipkin-Nogami obtained from the code HFBTHO \cite{mass_sly4}, produce
good agreement with the measured  $Q_\beta$ and $S_n$ energies.

In the upper panels of Fig. \ref{fig_qbeta_sn_red} we can see the 
experimental $Q_\beta$ values (black dots) \cite{audi12} for Zr (a) and 
Mo (b) isotopes. This information is only available for the isotopes 
$^{100-112}$Zr and $^{104-117}$Mo, although the values beyond $^{105}$Zr and 
$^{109}$Mo are evaluated from systematics. These values are compared with
the predictions of the two mass models mentioned above. In the lower
panels we have the neutron separation energies $S_n$ corresponding to 
the daughter isotopes of Zr and Mo, Nb (c) and Tc (d), respectively.
In this case the energies beyond  $^{109}$Nb and $^{113}$Tc are not 
directly measured, but deduced from systematics. In what follows the 
results for $T_{1/2}$ and $P_n$  will be calculated by using experimental 
$Q_\beta$ and $S_n$ values when available, and values from SLy4 and DZ-28 
in other cases.

In Figs. \ref{fig_t_zr_1} and \ref{fig_t_mo_1} we compare the measured 
$\beta$-decay half-lives  with the theoretical results obtained with 
the prolate, oblate, and spherical equilibrium shapes, for Zr and Mo 
isotopes, respectively. Three regions can be distinguished in these 
figures. A first region (I) with solid symbols for prolate, oblate, and 
spherical shapes, where experimental $Q_\beta$ values are used, 
$^{100-105}$Zr and $^{104-111}$Mo. A second region (II) with open symbols
that correspond to $^{106-112}$Zr and $^{112-117}$Mo, where we use 
$Q_\beta$ values extracted from systematics, from SLy4 and from DZ-28,
and a third region (III), $^{113-116}$Zr and $^{118-120}$Mo, where only
calculated $Q_\beta$ values are used.

In the first region (I) of both figures, the half-lives from prolate
shapes are always larger than those from oblate shapes and both of them
are larger than the half-lives from spherical nuclei.  This 
feature can be easily understood from Figs. 
\ref{fig_bgt_zr_qbeta}-\ref{fig_bgt_mo_qbeta}, where we can see how
the GT strength contained in the energy region below  $Q_\beta$ is 
much lower in the prolate case than in the oblate and spherical
cases. Comparison with the experiment in this region indicates that
the trend observed in Zr isotopes is nicely reproduced by the prolate
shapes with a tendency to overestimate the half-lives. In this region 
an oblate component that produces smaller half-lives would help in 
the agreement with experiment. In the case of the lighter Mo isotopes
in region (I) the prolate shapes overestimate clearly the measured 
half-lives, while the oblate shapes underestimate them. A shape 
coexistence in lighter Mo isotopes would produce in this case a better 
agreement with experiment.
In the middle region (II) we can see that the spread of the results
due to the nuclear deformation is comparable with the spread due
to the uncertainty in the $Q_\beta$ value. Nevertheless, we can still
perceive the tendency of the prolate (spherical) shapes to overestimate 
(underestimate) the half-lives with results from oblate shapes
somewhat in between. One should also keep in mind that the 
experimental half-lives for  $^{111,112}$Zr  and for $^{116-117}$Mo
are not properly measured values but extracted from systematics.
In the third region (III) of heavier isotopes we obtain spherical 
or close to sphericity shapes. The predicted half-lives corresponding 
to these shapes are relatively large and almost flat. This tendency 
seems to be at variance with the tendency shown by the experimental 
half-lives extracted from systematics that continue decreasing. 
Measurements in this region will be crucial to determine whether 
isotopes in this mass region become spherical as we approach the magic 
neutron number.

In the next two figures we simplify the comparison of various 
calculations for the half-lives by showing only the most relevant
ones. Thus, in Figs. \ref{fig_t_zr_2} and \ref{fig_t_mo_2} we 
compare the measured $\beta$-decay half-lives with the calculations 
corresponding only to the ground states of the isotopes. For Zr 
isotopes in Fig. \ref{fig_t_zr_2} we plot half-lives for prolate 
shapes for  $^{100-112}$Zr and for spherical shapes for $^{113-116}$Zr. 
Similarly, for Mo isotopes in Fig. \ref{fig_t_mo_2} we plot
half-lives for prolate shapes for $^{104-107}$Mo, for oblate shapes for  
$^{108-117}$Mo, and for spherical shapes for $^{118-120}$Mo.
Similar to the previous figures, one can also distinguish
three regions according to the $Q_\beta$ values used.
In Fig. \ref{fig_t_zr_2} for Zr isotopes we see again how the prolate
ground state shapes describe well the pattern of the half-lives.
Nevertheless the agreement can be improved by reducing somewhat
the half-lives of the lighter isotopes with an oblate component
predicted by the calculations. In the middle region, the experimental 
pattern is reasonably well reproduced by the calculations, but the SLy4 
masses reproduce better the measured half-lives. Finally for the heavier 
isotopes the predicted half-lives corresponding to spherical nuclei
are almost flat.
In Fig. \ref{fig_t_mo_2} for Mo isotopes, a shape coexistence of the 
prolate ground state shape with the excited oblate shape in the
lighter Mo isotopes would produce in a better agreement with experiment. 
In the middle region, the oblate shapes reproduce quite satisfactorily 
the measurements and no big differences are found between the
half-lives obtained by using SLy4 or experimental $Q_\beta$ values.
As in the case of Zr isotopes, the heavier spherical Mo isotopes 
predict relatively large half-lives that seems to deviate from the
decreasing exhibit by the experimental values from systematics.

Our results for the half-lives and GT strength distributions in 
Zr isotopes agree well with the calculations by Yoshida \cite{yoshida_13},
where similar results are presented for even-even Zr isotopes
within a self-consistent QRPA with Skyrme interactions, using only
the strengths of the pairing interaction as extra parameters and
a constant strength for the residual $T=0$ pairing interaction, which
is fitted to reproduce the half-life of $^{100}$Zr.
Our results agree also qualitatively with those in Ref. \cite{fang_13} 
for $T_{1/2}$ and $P_n$ in Zr and Mo isotopes, where QRPA calculations 
are performed using deformed Woods-Saxon potentials and
realistic CD-Bonn residual forces. First-forbidden transitions 
were also considered in \cite{fang_13}, although it was found on
the basis of the branching ratio for the forbidden decays
that in this mass region their effect can be neglected.

In the next figures (Figs. \ref{fig_pn_zr_1}-\ref{fig_pn_mo_2}),
we show the same type of information as in Figs. 
\ref{fig_t_zr_1}-\ref{fig_t_mo_2}, but for the $\beta$-delayed 
neutron-emission probabilities $P_n$, expressed as a percentage. 
Experimental $P_n$ values \cite{pereira} are only upper limits except 
for the cases $^{110,111}$Mo. In all cases the calculations underestimate
the existing measurements. This means that in the range of isotopes where
these measurements exist, the relative GT strength contained in the energy 
region below (above) $S_n$ is overestimated (underestimated) theoretically.
The general isotopic trend observed is an increase of $P_n$ with $A$,
obviously related to the decrease of $S_n$ as we approach the neutron
drip line. The behavior of $P_n$ for spherical shapes is somewhat special.
It appears clearly below the deformed cases up to $A\sim 112$ 
($A\sim 116$) in Zr (Mo) isotopes, and then suddenly it increases up
to practically $P_n=100\%$ in the heavier isotopes.
The origin of this behavior can be traced back to the profile of the
accumulated strength distributions in spherical nuclei showing a sharp 
increase in Figs. \ref{fig_bgt_zr}-\ref{fig_bgt_mo} at variance with the 
smooth distribution of the deformed cases denoting the high fragmentation 
of the strength.
It is also worth mentioning the extreme sensitivity of $P_n$ to  $S_n$
in the case of spherical nuclei. Since most of the strength in spherical 
nuclei is very localized in energy, $P_n$ can change drastically if this
strength is located a little bit above or a little bit below $S_n$. 
Thus, we can see for example how $P_n$ in $^{112}$Zr changes from $0.3\%$ 
to $100\%$ when we use $S_n$ from DZ-28 (dotted lines) or from SLy4 
(dashed lines). The $S_n$ values for $^{112}$Nb change from 2.92 MeV with 
SLy4 to 3.31 MeV for DZ-28 (Fig.\ref{fig_qbeta_sn_red}(c)), values that 
are respectively below and above the energy where the GT strength peaks in  
$^{112}$Zr (Fig.\ref{fig_bgt_zr_qbeta}).
Similarly, the sharp change in $P_n$ between spherical $^{112}$Zr and 
$^{113}$Zr isotopes is related to the relative position of the strength 
with respect to $S_n$, which appears below $S_n$ in $^{112}$Zr and above it 
in  $^{113}$Zr (see Fig. \ref{fig_bgt_zr_qbeta}).

Similar general comments are valid for Mo isotopes. In this case
particular behaviors in the spherical case can be understood as
well by means of the interplay between $S_n$ values and the energies
where the GT strength is peaked.
This is the case of $^{118}$Mo, where the low $P_n$ value found
for DZ-28 is correlated to the high value of $S_n$ in 
Fig. \ref{fig_qbeta_sn_red}(d). This is also the case of $^{119}$Mo, 
where the low $P_n$ value for SLy4 is due to the large value of $S_n$.
We can also mention the strange behavior of $^{116}$Mo in the prolate 
cases that decrease suddenly up to values similar to the spherical ones.
This is related to the fact that the prolate deformation in this isotope
has collapsed almost to a spherical shape as it can be seen in 
Fig. \ref{fig_beta}.

Finally, we can see in Figs. \ref{fig_pn_zr_2}- \ref{fig_pn_mo_2}
the percentage $P_n$ values, but showing only the most relevant 
information that considers only the ground state shapes, as it has 
been done in Figs. \ref{fig_t_zr_2}- \ref{fig_t_mo_2}. 
The general behavior discussed above can be seen 
more clearly in these simplified plots with increasing values for heavier 
isotopes that saturates to a $100\%$ value when the spherical shape becomes
the ground state in the heavier isotopes with the particularity in the
case of $^{119}$Mo discussed above.

\section{CONCLUSIONS}

A microscopic nuclear approach based on a deformed QRPA calculation on 
top of a self-consistent mean field obtained with the SLy4 Skyrme 
interaction has been used to study the decay properties of 
neutron-rich Zr and Mo isotopes. The nuclear model and interaction
have been successfully tested in the past providing good agreement 
with the available experimental information on bulk properties all 
along the nuclear chart. 
Decay properties in other mass regions have been well reproduced as well.
In this work we have studied even and odd isotopes, encompassing the 
whole isotopic chains $^{100-116}$Zr and $^{104-120}$Mo. 
The structural evolution in these isotopes has been studied 
from their PECs. We have found competing oblate and prolate shapes
in the lighter isotopes and spherical shapes becoming ground states
in the heavier. Then, Gamow-Teller strength distributions, 
$\beta$-decay half-lives, and $\beta$-delayed neutron emission
probabilities corresponding to the equilibrium shapes of the 
respective isotopes have been computed.

The isotopic evolution of the GT strength distributions exhibit some
typical features, such as the GT resonances that increase in energy
and strength as the number of neutrons increase. Effects of deformation
are hard to see on a global scale, but they become apparent in the 
low excitation energy below $Q_\beta$ energies, a region that determines
the half-lives. Half-lives have been calculated with experimental values
of $Q_\beta$ when available and with values extracted from mass models 
otherwise. The spread of the results due to the nuclear shape and
to the lack of knowledge of $Q_\beta$ is comparable. Thus, it is
very important to continue measuring the masses of exotic nuclei
to avoid these type of uncertainties.
The half-lives are in general well described in the lighter and medium
isotopes by using the ground state shapes, although they would 
improve a bit with some admixtures of the isomeric shapes.
In this region, the half-lives obtained with spherical shapes are
clearly below the experiment and the results from deformed shapes.
For heavier isotopes where the half-lives have not been measured yet,
the calculations predict a rather flat behavior corresponding to
spherical ground states.
$P_n$ values are not well reproduced. The calculations underestimate 
them although only upper limits are measured in most cases. Nevertheless,
the general trend observed is followed when using deformed shapes, 
whereas spherical shapes produce extremely low $P_n$ values. The 
scenario changes suddenly in the heavier isotopes, where spherical 
shapes produce $P_n$ values of $100\%$.

Thus, experimental information on the energy distribution of the GT strength 
is a valuable piece of knowledge about nuclear structure in this mass region. 
This is within the present perspectives in the case of the lighter isotopes 
considered in this work.
Similarly, measuring the half-lives of the heavier isotopes will be highly 
beneficial to model reliably the r process and to constrain theoretical 
nuclear models. This possibility is also open within present capabilities
at Riken.

\begin{acknowledgments}
This work was supported by Ministerio de Econom\'\i a y Competividad
(Spain) under Contracts No. FIS2011--23565 and FPA 2011--24553 and the 
Consolider-Ingenio 2010 Programs CPAN CSD2007-00042.
It was also supported in part by the Joint Institute for Nuclear Astrophysics 
(JINA) under NSF Grant No. PHY-02-16783 and the National Superconducting 
Cyclotron Laboratory (NSCL), under NSF Grant No. PHY-01-10253.
\end{acknowledgments}


\begin{thebibliography}{00}


\bibitem{wood92} J. L. Wood, K. Heyde, W. Nazarewicz, M. Huyse, and P.
Van Duppen, Phys. Rep. {\bf 215}, 101 (1992).
\bibitem{heyde11} K. Heyde and  J. L. Wood, Rev. Mod. Phys. {\bf 83}, 1467 (2011).
\bibitem{xiang_12} J. Xiang, Z. P. Li, Z. X. Li, J. M. Yao, and J. Meng,
 Nucl. Phys. {\bf A873}, 1 (2012).
\bibitem{mei_12} H. Mei, J. Xiang, J. M. Yao, Z. P. Li, and J. Meng,
 Phys. Rev. C {\bf 85}, 034321 (2012).
\bibitem{bonche85} P. Bonche, H. Flocard, P.-H. Heenen, S. J. Krieger, and
M. S. Weiss, Nucl. Phys. {\bf A443}, 39 (1985).
\bibitem{bender08} M. Bender, G. F. Bertsch, and P.-H. Heenen, 
Phys. Rev. C {\bf 78}, 054312 (2008).
\bibitem{bender09} M. Bender, K. Bennaceur, T. Duguet, P.-H. Heenen,
T. Lesinski, and J. Meyer, Phys. Rev. C {\bf 80}, 064302 (2009).
\bibitem{rayner1} R. Rodriguez-Guzman, P. Sarriguren, L. M. Robledo, and 
S. Perez-Martin,   Phys. Lett. {\bf B 691}, 202 (2010).
\bibitem{rayner2} R. Rodriguez-Guzman, P. Sarriguren, and L. M. Robledo,
Phys. Rev. C {\bf 82}, 044318 (2010); Phys. Rev. C {\bf 83}, 044307 (2011).
\bibitem{sumikama_11} T. Sumikama  {\it et al.},  Phys. Rev. Lett. {\bf 106},
202501 (2011).
\bibitem{mach} H. Mach  {\it et al.}, Phys. Lett. {\bf B 230}, 21
(1989); Phys. Rev.  C {\bf 41}, 350 (1990).
\bibitem{goodin} C. Goodin  {\it et al.}, Nucl. Phys. {\bf A787},
231c (2007).
\bibitem{urban} W. Urban  {\it et al.}, Nucl. Phys. {\bf A689}, 605 (2001).
\bibitem{campbell} P. Campbell {\it et al.}, Phys. Rev. Lett. {\bf 89}, 
082501 (2002).
\bibitem{charlwood} F. C. Charlwood {\it et al.}, Phys. Lett. {\bf B 674},
23 (2009).
\bibitem{bbhf} E. M. Burbidge, G. M. Burbidge, W. A. Fowler, and
F. Hoyle, Rev. Mod. Phys. {\bf 29}, 547 (1959).
\bibitem{cowan91} J. J. Cowan, F.-K. Thielemann, and J. W. Truran,
Phys. Rep. {\bf 208}, 267 (1991).
\bibitem{kra93} K.-L.~Kratz, J.-P.~Bitouzet, F.-K.~Thielemann, 
P.~M\"oller, and B.~Pfeiffer,  Ap.~J. {\bf 403}, 216 (1993).
\bibitem{jyvaskyla} http://research.jyu.fi/igisol/JYFLTRAP\_masses/
\bibitem{pereira} J. Pereira {\it et al.},  Phys. Rev. C {\bf 79},
035806 (2009).
\bibitem{nishimura_11} S. Nishimura {\it et al.},  Phys. Rev. Lett. {\bf 106},
052502 (2011).
\bibitem{pfeiffer01} B. Pfeiffer,  K.-L. Kratz, F.-K. Thielemann, and
W. B. Walters,  Nucl. Phys. {\bf A693}, 282 (2001).
\bibitem{sun08} B. Sun, F. Montes, L. S. Geng, H. Geissel, Yu. A. Litvinov,
and J. Meng,  Phys. Rev. C {\bf 78}, 025806 (2008).
\bibitem{doba96} J. Dobaczewski, W. Nazarewicz, T. R. Werner, J. F. Berger,
C. R. Chinn, and J. Decharg\'e, Phys. Rev. C {\bf 53}, 2809 (1996).
\bibitem{ETFSIQ} J. M. Pearson, R. C. Nayak, and S. Goriely,
Phys. Lett. {\bf B 387}, 455 (1996).
\bibitem{engel} J. Engel, M. Bender, J. Dobaczewski, W. Nazarewicz,
and R. Surman,  Phys. Rev. C {\bf 60}, 014302 (1999).
\bibitem{borzov3} I. N. Borzov, J. J. Cuenca-Garc\'{\i}a, K. Langanke,
G. Mart\'{\i}nez-Pinedo, and F. Montes, Nucl. Phys. {\bf A814}, 159 (2008).
\bibitem{niksic2005} T. Niksic, T. Marketin, D. Vretenar, N. Paar,
and P. Ring, Phys. Rev. C {\bf 71}, 014308 (2005).
\bibitem{moller1} J. Krumlinde and P. M\"oller, Nucl. Phys.
{\bf A417}, 419 (1984).
\bibitem{moller2} P. M\"oller and J. Randrup, Nucl. Phys.
{\bf A514}, 1 (1990).
\bibitem{moller3} P. M\"oller, B. Pfeiffer, and K.-L. Kratz, Phys. Rev. C 
{\bf 67}, 055802 (2003).
\bibitem{moller08} P. M\"oller, R. Bengtsson, B. G. Carlsson, P. Olivius,
T. Ichikawa, H. Sagawa, and A. Iwamoto, At. Data Nucl. Data Tables {\bf 94},
758 (2008).
\bibitem{homma} H. Homma, E. Bender, M. Hirsch, K. Muto, H. V.
Klapdor-Kleingrothaus, and T. Oda, Phys. Rev. C {\bf 54}, 2972 (1996).
\bibitem{hir1}  M. Hirsch, A. Staudt, K. Muto, and H. V.
Klapdor-Kleingrothaus, Nucl. Phys. {\bf A535}, 62 (1991).
\bibitem{hir2} K. Muto, E. Bender, T. Oda, and H. V. Klapdor-Kleingrothaus,
Z. Phys.  {\bf A 341}, 407 (1992).
\bibitem{hamamoto}   F. Frisk, I. Hamamoto, and X.Z. Zhang, Phys. Rev. C 
{\bf 52}, 2468 (1995).
\bibitem{sarri1}  P. Sarriguren, E. Moya de Guerra, A. Escuderos, and
A. C. Carrizo, Nucl. Phys. {\bf A635}, 55 (1998).
\bibitem{sarri2} P. Sarriguren, E. Moya
de Guerra, and A. Escuderos, Nucl. Phys. {\bf A691},  631 (2001).
\bibitem{sarri_pere} P. Sarriguren and J. Pereira,  Phys. Rev. C {\bf 81}, 
064314 (2010).
\bibitem{audi12}  G. Audi {\it et al.}, Chinese Physics C {\bf 36}, 1157 (2012);
 1603 (2012).
\bibitem{sly4} E. Chabanat, P. Bonche, P. Haensel, J. Meyer, and
R. Schaeffer, Nucl. Phys. {\bf A635}, 231 (1998).
\bibitem{stoitsov} M. V. Stoitsov, J. Dobaczewski, W. Nazarewicz, S. Pittel, 
and D. J. Dean, Phys. Rev. C {\bf 68}, 054312 (2003).
\bibitem{vautherin} D. Vautherin and D. M. Brink, Phys. Rev. C
{\bf 5}, 626 (1972); D. Vautherin, Phys. Rev. C {\bf 7}, 296 (1973).
\bibitem{constraint} H. Flocard, P. Quentin, A. K. Kerman, and D.
 Vautherin, Nucl. Phys. {\bf A203}, 433 (1973).
\bibitem{sarri3} P. Sarriguren, E. Moya de Guerra, and A. Escuderos,
Nucl. Phys. {\bf A658}, 13 (1999).
\bibitem{sarri4} P. Sarriguren, E. Moya de Guerra, and A. Escuderos,
Phys. Rev. C {\bf 64}, 064306 (2001).
\bibitem{sarri_wp} P. Sarriguren, R. Alvarez-Rodr\'{\i}guez, and E.
Moya de Guerra, Eur. Phys. J. A {\bf 24}, 193 (2005).
\bibitem{bm} A. Bohr and B. Mottelson, {\em Nuclear Structure},
Vols. I and II, (Benjamin, New York 1975).
\bibitem{gove} N. B. Gove and M. J. Martin, Nucl. Data Tables {\bf 10},
205 (1971).
\bibitem{skalski97} J. Skalski, S. Mizutori, and W. Nazarewicz,
Nucl. Phys. {\bf A617}, 281 (1997).
\bibitem{FRDM} P. M\"oller, J. R. Nix, W. D. Myers, and W. J.
Swiatecki, At. Data Nucl. Data Tables {\bf 59}, 185 (1995).
\bibitem{lala2} G. A. Lalazissis, S. Raman, and P. Ring, At. Data Nucl.
Data Tables {\bf 71}, 1 (1999).
\bibitem{hilaire} S. Hilaire and M. Girod, Eur. Phys. J. A {\bf 33},
237 (2007); http://www-phynu.cea.fr/.
\bibitem{moya} E. Moya de Guerra, Phys. Rep. {\bf 138}, 293 (1986).
\bibitem{raman} S. Raman, C. W. Nestor, JR., and P. Tikkanen,
At. Data Nucl. Data Tables {\bf 78}, 1 (2001).
\bibitem{stone} N. J. Stone, At. Data Nucl. Data Tables {\bf 90}, 75 (2005).
\bibitem{ensdf} Evaluated Nuclear Structure Data File (ENSDF),
http://www.nndc.bnl.gov/ensdf/
\bibitem{exp_poirier} E. Poirier {\it et al.}, Phys. Rev. C {\bf 69},
034307 (2004).
\bibitem{exp_nacher}  E. N\'acher {\it et al.}, Phys. Rev. Lett. {\bf 92},
232501 (2004).
\bibitem{jokinen} A. Jokinen, T. Enqvist, P. P. Jauho, M. Leino,
J. M. Parmonen, H. Penttil\"a, J. \"Ayst\"o,  and  K. Eskola,
Nucl. Phys. {\bf A584}, 489 (1995).
\bibitem{wang04} J. C. Wang  {\it et al.}, Eur. J. A {\bf 19}, 83 (2004).
\bibitem{rinta07} S. Rinta-Antila {\it et al.}, Eur. J. A {\bf 31}, 1 (2007).
\bibitem{langanke} K. Langanke and G. Martinez-Pinedo,
Nucl. Phys. {\bf A673}, 481 (2000).
\bibitem{sarri09_1} P. Sarriguren, Phys. Rev. C {\bf 79}, 044315 (2009).
\bibitem{sarri09_2} P. Sarriguren, Phys. Lett. {\bf B 680}, 438 (2009).
\bibitem{DufloZuker} J. Duflo and A. P. Zuker, Phys. Rev. C {\bf 52},
R23 (1995).
\bibitem{mass_sly4} M. V. Stoitsov, J. Dobaczewski, W. Nazarewicz, and
P. Ring, Comp. Phys. Comm. {\bf 167}, 43 (2005).
\bibitem{yoshida_13} K. Yoshida, Prog. Theor. Exp. Phys. 113D02 (2013).
\bibitem{fang_13} D. L. Fang, B. A. Brown, and T. Suzuki, 
 Phys. Rev. C {\bf 88}, 024314 (2013).


\end{thebibliography}
\end{document}